\documentclass[aps,print,prb,twocolumn,superscriptaddress]{revtex4-2}
\usepackage[hidelinks]{hyperref}
\hypersetup{
	colorlinks,
	linkcolor={red},
	citecolor={blue},
	urlcolor={blue}
}
\usepackage{balance}
\usepackage{amsmath, amssymb}
\usepackage{suffix}
\usepackage{mathtools}
\usepackage[utf8]{inputenc}
\usepackage{booktabs}
\usepackage{cases}
\usepackage[multiple]{footmisc}
\usepackage{dcolumn}
\usepackage{color,soul}
\usepackage{rotating}
\usepackage{perpage}
\usepackage{siunitx}
\usepackage{xcolor}
\usepackage{amsmath}
\usepackage{tikz}
\usepackage[T1]{fontenc}
\usepackage{etoolbox}
\usepackage{graphicx}
\usepackage{float}
\usepackage{collref}
\usepackage{multirow}
\usepackage{bm}
\usepackage{url}

\usepackage{tikz-3dplot}
\usepackage{accents}

\makeatletter
\newcommand{\doublewidetilde}[1]{{%
		\mathpalette\double@widetilde{#1}%
}}
\newcommand{\double@widetilde}[2]{%
	\sbox\z@{$\m@th#1\widetilde{#2}$}%
	\ht\z@=.9\ht\z@
	\widetilde{\box\z@}%
}
\makeatother

\begin{document}

\title{Strain-stabilized altermagnetism and conductivity anisotropy in FeSb$_2$}

\author{Masoumeh Davoudiniya} 
\email{md2102@georgetown.edu}
\affiliation{Department of Physics, Georgetown University, Washington, D.C. 20057, USA}

\author{Alyssa M. Kennedy}
%\email{}
\affiliation{Department of Physics, Georgetown University, Washington, D.C. 20057, USA}
\affiliation{Department of Chemistry and Biochemistry, Denison University, Granville, OH 43023, USA}

\author{Amy Y. Liu}
\email{liua@georgetown.edu}
\affiliation{Department of Physics, Georgetown University, Washington, D.C. 20057, USA}

\author{Gen Yin}
\email{gen.yin@georgetown.edu}
\affiliation{Department of Physics, Georgetown University, Washington, D.C. 20057, USA}

\date{\today}

\begin{abstract}
We show that FeSb$_2$ experiences a transition from a conventional antiferromagnet to an altermagnet when tensile strain is applied. 
In the altermagnetic phase, the lifted Kramers degeneracy results in  spin splitting up to $\sim0.2\thinspace\textrm{eV}$ near the Fermi level even without spin-orbit coupling. 
The transition to the altermagnetic phase is accompanied by a dramatic change in the Fermi-surface geometry, which leads to a uniaxial conductivity anisotropy up to $\sim60\%$, much greater than those observed in typical ferromagnetic metals. 
Using density-functional theory and Wannier interpolated Fermi surfaces, we show that this magnetotransport behavior may function as an experimental indicator of the transition to the altermagnetic phase. 
These findings highlight FeSb$_2$ as a versatile, strain-tunable platform for exploring and utilizing altermagnetic transport phenomena for spintronic devices.

\end{abstract}

\maketitle

\section{Introduction}

Magnetic order is commonly classified by the arrangement of local moments and the resulting net magnetization~\cite{PhysRevX.12.040501,PhysRevX.12.040002,PhysRevX.12.031042,doi:10.7566/JPSJ.88.123702,PhysRevX.12.011028,PhysRevB.102.014422,PhysRevB.99.184432,PhysRevMaterials.5.014409,he2025altermagnetismttprimedeltafermihubbardmodel}.

In many conventional collinear antiferromagnets (AFMs), antiunitary symmetries --- most commonly $\mathcal{PT}$, and in bipartite lattices, $\mathcal{T}\tau$, where $\mathcal{T}$ is time reversal, $\mathcal{P}$ is spatial inversion, and $\tau$ is a lattice translation --- enforce a twofold Kramers-like degeneracy of the electronic bands at each $\mathbf{k}$~\cite{Jungwirth2016,RevModPhys.90.015005}.
 Altermagnets (AMs) evade this constraint while remaining collinear and magnetically compensated \cite{PhysRevX.12.040501,PhysRevX.12.031042}. 
 Their electronic structure can therefore exhibit pronounced momentum-dependent spin splitting. 
 In strong AMs, the splitting is governed by the exchange field together with magnetic crystal symmetry rather than by spin-orbit coupling (SOC)~\cite{PhysRevX.12.031042}. 
 This combination of magnetic compensation and spin-split bands makes AMs attractive for information processing without a net magnetization~\cite{Jungwirth2016,mejkal2022}, potentially providing ultrafast switching dynamics and immunity to external magnetic fields for future spintronic devices.

Recent experiments have captured direct signatures of AMs. 
Photoemission measurements have resolved the altermagnetic lifting of the spin degeneracy in prototype systems~\cite{Krempaský2024,Reimers2024}.
Early experimental and theoretical efforts have also heavily focused on metallic candidates like RuO$_2$, where spectroscopic signatures of the broken time-reversal symmetry~\cite{Fedchenko2024} and anomalous transport responses~\cite{Feng2022} have been investigated.
In parallel, high-throughput materials-discovery efforts have rapidly expanded the list of proposed altermagnetic compounds across a wide variety of crystal classes and electronic properties~\cite{PhysRevMaterials.9.064403}. 
These developments shift the central question from merely identifying AMs to controlling the altermagnetic order and its transport consequences. 
Although ideal collinear AMs are fully compensated in equilibrium, this compensation can be modified by perturbations lowering the symmetry. 
Diverse external drives provide versatile routes to modulate the spin order, spin polarization, and transport responses~\cite{golub2025spinorientationelectriccurrent,Yarmohammadi2025Spin,Yarmohammadi2025SlowPhonon,Yarmohammadi2025Cavity,Yarmohammadi2026FloquetAMR}. 
In equilibrium, lattice distortions, strain, and reduced coordination at edges or interfaces can induce canting, piezomagnetism, or boundary magnetization. 
These structural perturbations are known to be important for many altermagnetic systems such as RuF$_4$~\cite{Milivojevi2024}, MnTe~\cite{Aoyama2024MnTe,chen_large_2026}, altermagnetic interfaces~\cite{Hodt2024} and piezoelectric structures~\cite{Yershov2024,Ogawa2025}. 
In thin films, strain is a particularly attractive equilibrium handle. 
It offers a highly tunable and accessible means to control the magnetic configuration by modifying the local bonding geometry and crystal field~\cite{Rondinelli2011}.

FeSb$_2$ is a promising platform to host strain-controlled altermagnetic physics.
While it has primarily been studied for its large thermopower~\cite{Bentien2007,PhysRevLett.114.236603,Takahashi2016}, the material also exhibits interesting electronic and magnetic properties.  
It is a correlated small-gap semiconductor~\cite{PhysRevB.72.045103,Xu2020} with narrow low-energy bands that can be sensitive to lattice deformations. 
In addition, spectroscopic and theoretical studies have captured a temperature-driven transition from a low-spin insulating state to a metallic high-spin state~\cite{Li2024}, underscoring the sensitivity of these narrow low-energy bands to changes in the local crystal field. 
Density functional theory (DFT) calculations have suggested that while the ground state of stoichiometric FeSb$_2$ is a conventional Kramers-degenerate AFM, moderate Cr or Co doping could drive the system into an altermagnetic ground state~\cite{Mazin2021}. 
Indeed, recent experiments by Shawon et al. demonstrated that below 3.5 K, the spin-compensated bulk magnetic order in Fe$_{0.85}$Cr$_{0.15}$Sb$_2$ generates a spontaneous anomalous Hall response~\cite{Shawon2026CrDopedFeSb2}. 
These observations provide support for altermagnetism in FeSb$_2$-based systems, motivating the search for a robust, continuous tuning knob to drive and monitor the transition between the AFM and AM phases.

In this work, we use DFT calculations to explore the effect of mechanical strain on the stability of different magnetic phases of stoichiometric FeSb$_2$. We find that tensile strain stabilizes the AM phase compared to the AFM ground state. 
Despite the orthorhombic lattice structure, the conductivity tensor for both AFM configurations considered is only weakly anisotropic.
On the other hand, the AM spin configuration 
induces a highly anisotropic conductivity tensor. We thus propose changes in anisotropy in electrical conductivity measurements as an indicator of the transition between the two types of magnetically compensated states, AFM and AM, in FeSb$_2$.

\begin{figure*}[t]
    \centering
    \includegraphics[width=0.49\linewidth]{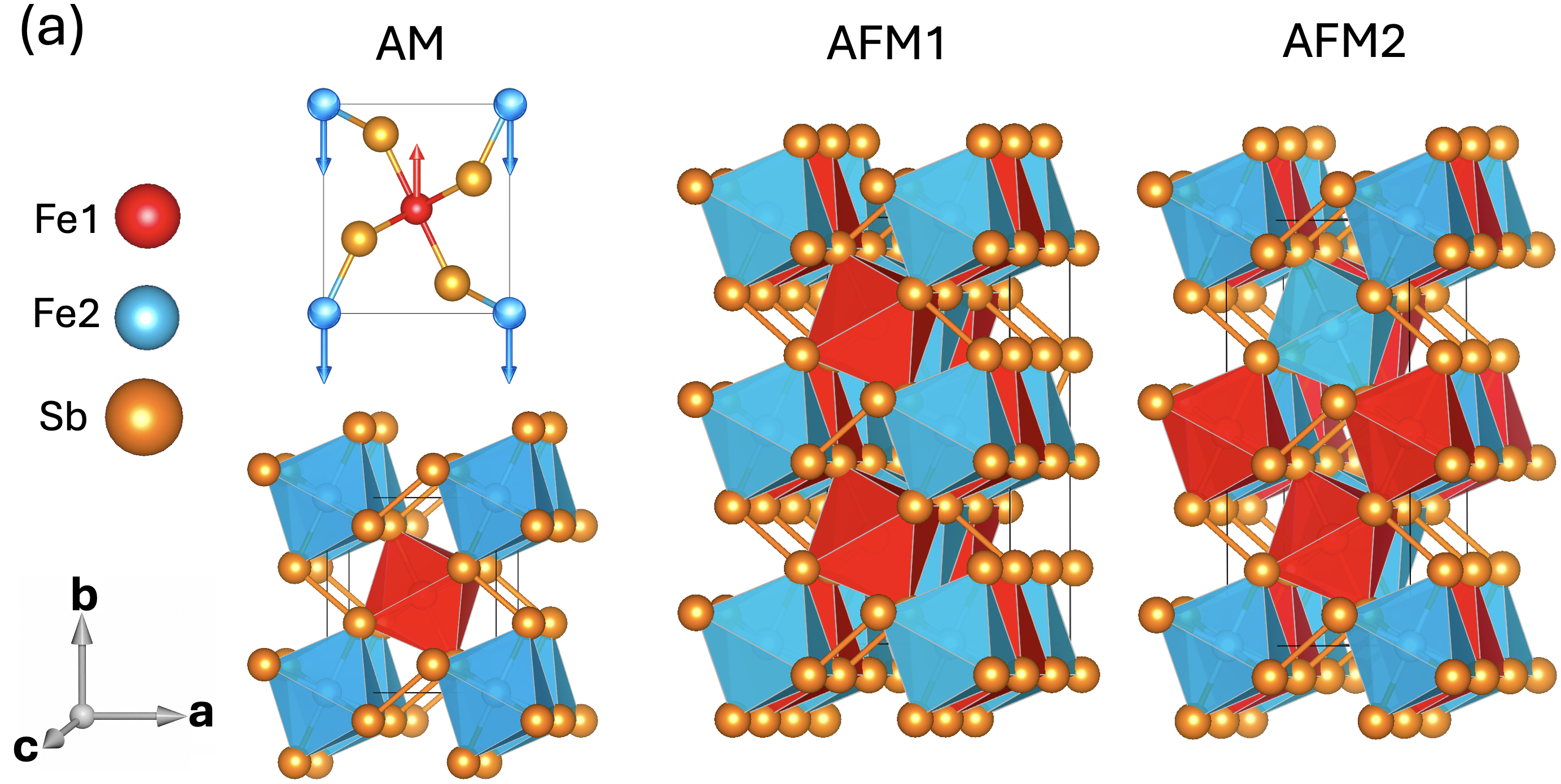} \hspace{0.2cm}
     \includegraphics[width=0.23\linewidth, trim={0cm -0.6cm 0cm 0cm}, clip]{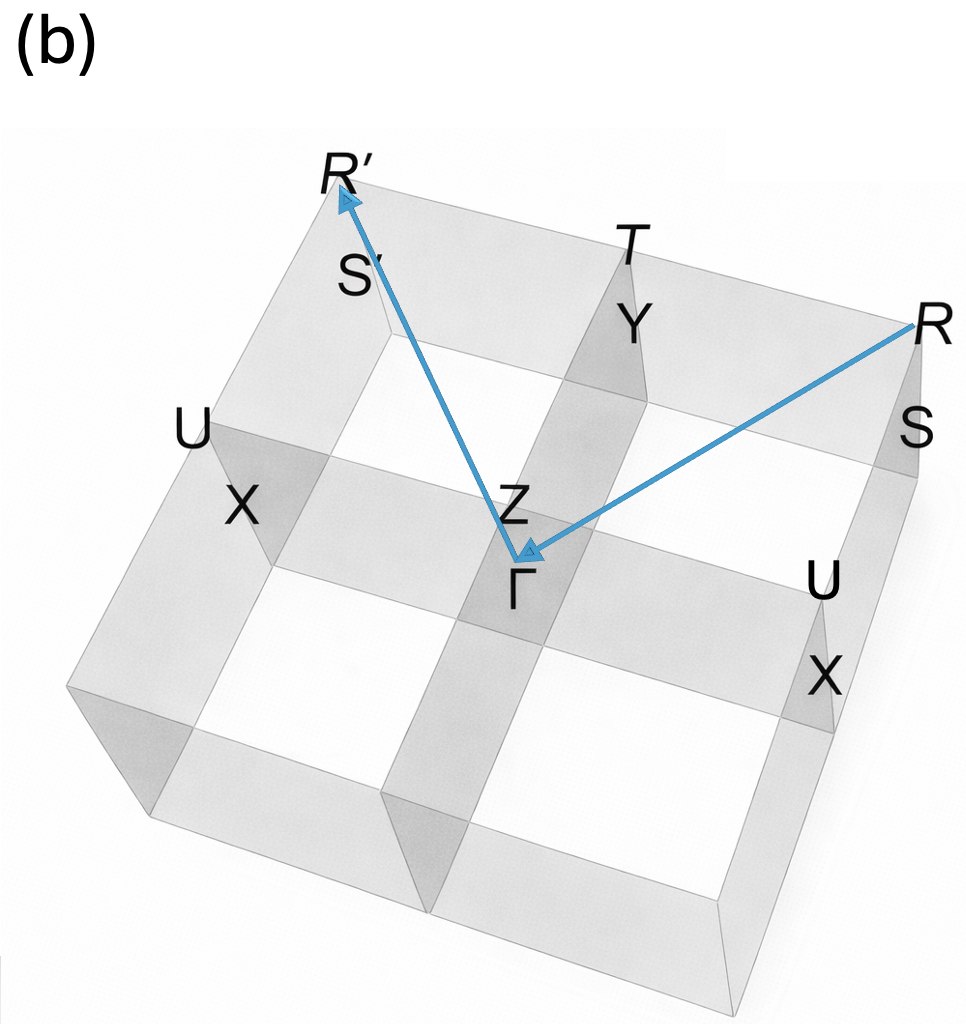}\hspace{0.2cm}\includegraphics[width=0.23\linewidth, trim={0cm -1.8cm 0cm 0cm}, clip]{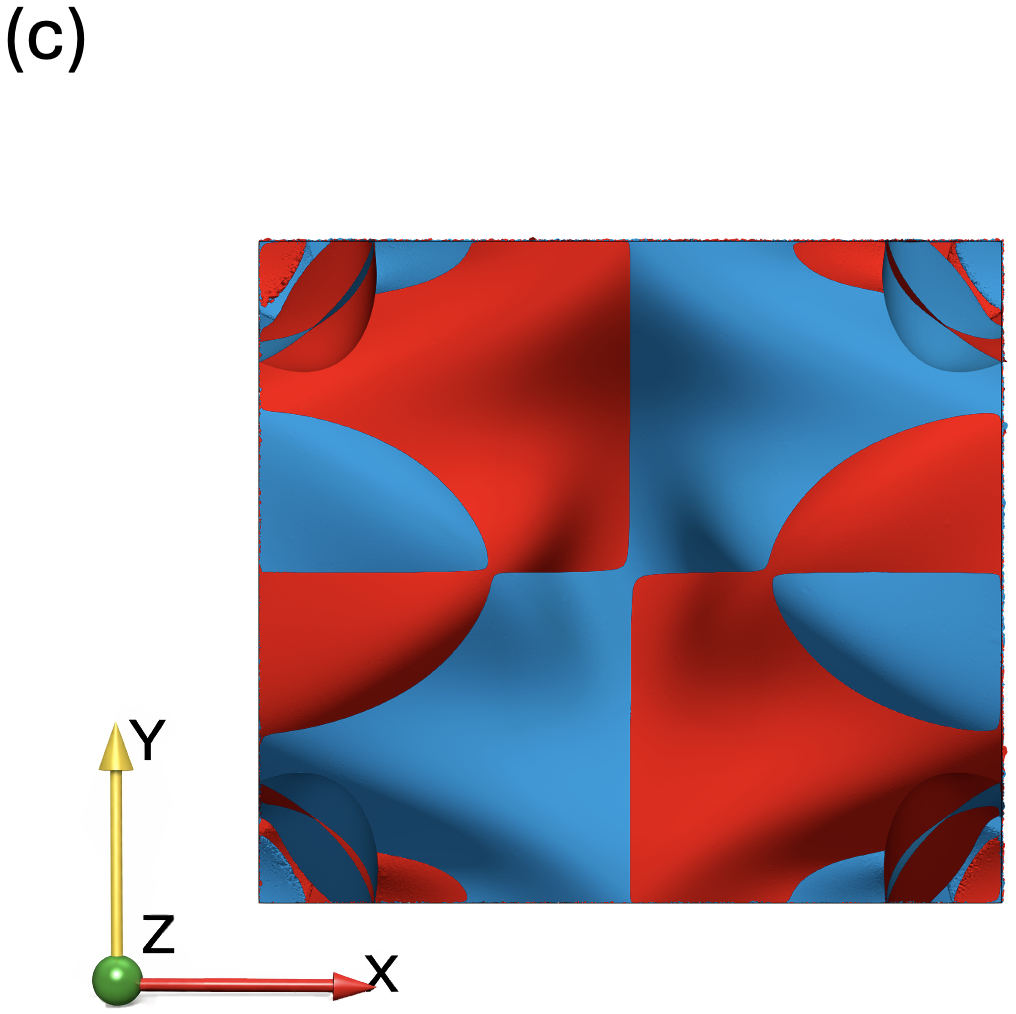}\\
\includegraphics[width=0.33\linewidth]{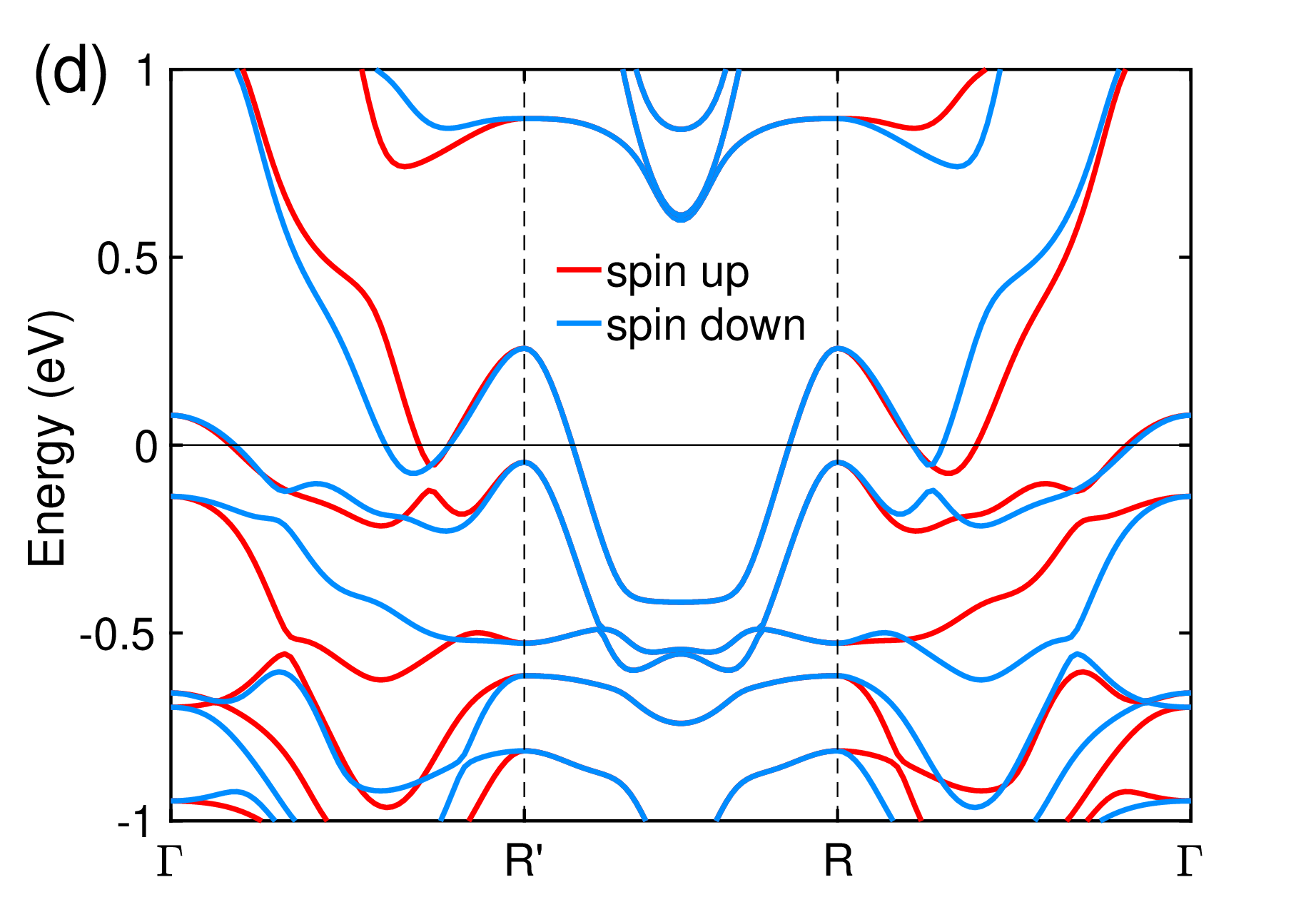}
\includegraphics[width=0.33\linewidth]{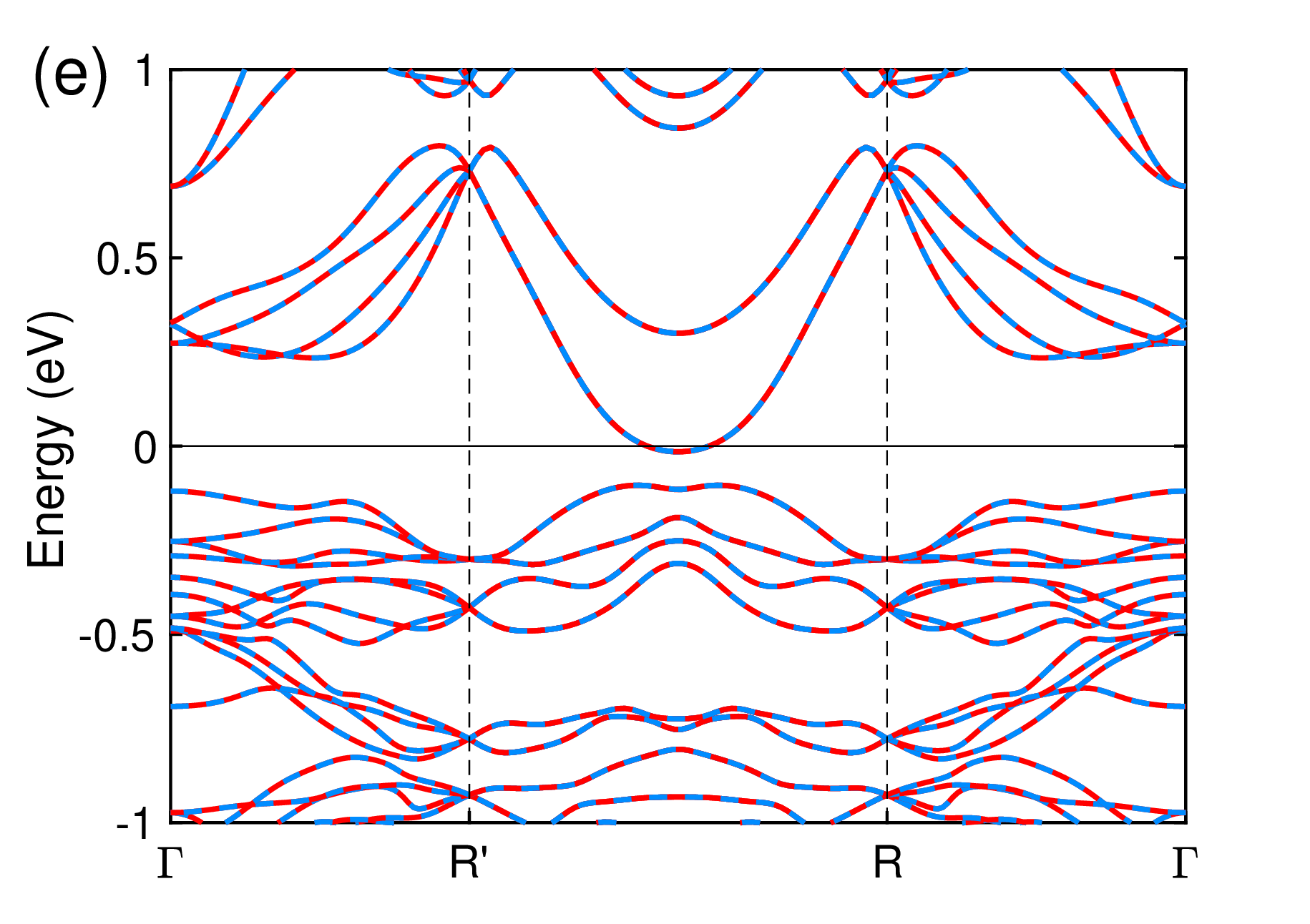}\includegraphics[width=0.33\linewidth]{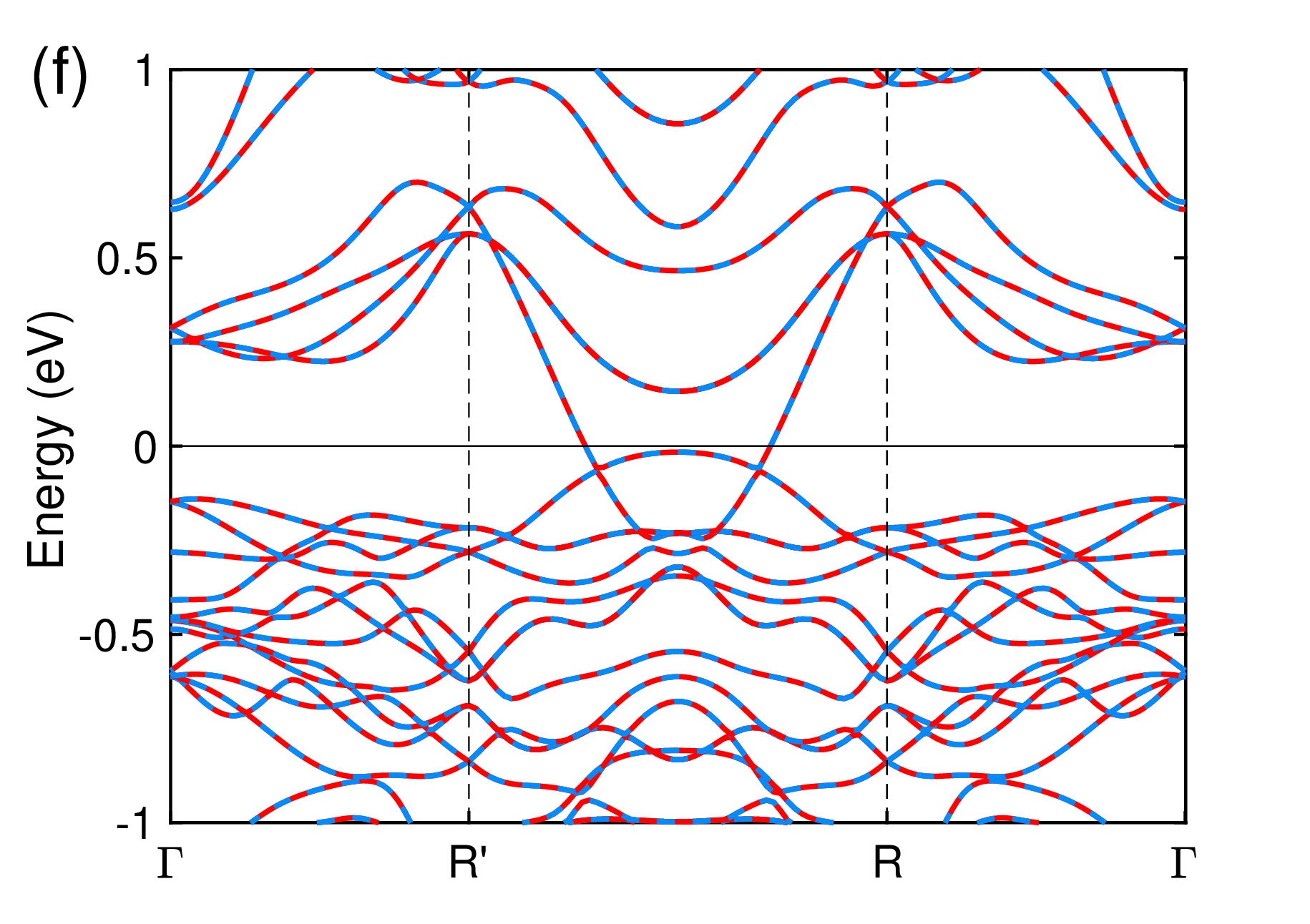}\\
\caption{Magnetic configurations and spin-resolved electronic structures of FeSb$_2$. (a) Magnetically compensated altermagnetic (AM) state and two conventional antiferromagnetic states (AFM1 and AFM2). For AM, the upper diagram shows the magnetic structure projected onto the $ab$ plane, with arrows indicating the local magnetic moments, while the lower diagram shows a three-dimensional polyhedral view highlighting the octahedral coordination of Fe by Sb.  (b) Brillouin zone of the orthorhombic reference cell of AM FeSb$_2$. For AFM1 and AFM2, the $R$ and $R'$ labels in panels (e) and (f) denote the analogous points in the magnetic Brillouin zones of the corresponding enlarged antiferromagnetic cells. % 
(c) Spin-polarized Fermi-surface map of the AM state. Red and blue denote opposite spin projections, 
showing the alternating spin-polarization pattern in reciprocal space. (d)--(f) Spin-resolved band structures of the AM, AFM1, and AFM2 states, respectively. The AM configuration exhibits clear momentum-dependent spin splitting despite vanishing net magnetization, whereas AFM1 and AFM2 remain spin-degenerate along the chosen path.  
}
    \label{fig1}
\end{figure*}

\section{Methodology}\label{method}

DFT calculations were performed using the projector-augmented-wave (PAW) method as implemented in \textsc{VASP}~\cite{Kresse1996CMS,Kresse1996PRB,Blochl1994,KresseJoubert1999}, with exchange--correlation treated within the PBE generalized-gradient approximation~\cite{PBE1996}.
Following the FeSb$_2$ altermagnetism study of Mazin \textit{et al.}\cite{Mazin2021}, we chose not to apply a static Hubbard $U$ correction, since the correction  
may alter the ground-state magnetic configuration, and may also overestimate the semiconducting gap in some implementations\cite{Lukoyanov2006FeSb2,Kuhn2013CrSb2FeSb2,Kang2018FeSb2Analogs}. 
Accordingly, the present work should be understood as a PBE-level comparison of strain-dependent magnetic energetics and its low-temperature transport consequences. 
A quantitative treatment of many-body gap renormalization and dynamical correlation effects lies beyond the scope of this study \cite{Tomczak2010ThermopowerFeSb2}.

The production calculations used a plane-wave cutoff of 520~eV, and electronic self-consistency was converged to $10^{-6}$~eV.
Structures were relaxed until residual forces were below $5\times10^{-3}$~eV\,\AA$^{-1}$.
Brillouin-zone integrations employed $\Gamma$-centered Monkhorst--Pack meshes \cite{MonkhorstPack1976}.
For the primitive marcasite FeSb$_2$ cell, which is orthorhombic and contains two formula units, a $7\times6\times12$ mesh of {\bf k} points was used. Equivalent {\bf k}-point meshes were used for supercells representing different magnetic configurations.  
Unless stated otherwise, calculations were performed at the scalar relativistic level. Fully relativistic calculations including the spin-orbit coupling (SOC) interaction were performed in some test cases.

For efficient calculation of transport properties, the DFT electronic band structures were fit to a tight-binding model.  
Maximally localized Wannier functions were constructed with \textsc{Wannier90}~\cite{Mostofi2008,Mostofi2014,Pizzi2020} using Fe $3d$ and Sb $p$ projections~\cite{MarzariVanderbilt1997,Souza2001}. The Wannier-interpolated bands were verified to reproduce the DFT band structure within the chosen energy window, and the resulting tight-binding Hamiltonians were used for subsequent analysis with the \textsc{QuantumGeom}~\cite{Yu2021PRB,chen_large_2026} package.

We evaluate the semiclassical longitudinal conductivity within the constant-relaxation-time approximation (CRTA)~\cite{ZimanBook,Madsen2006BoltzTraP}, assuming a common, {\bf k}-independent relaxation time $\tau$. 
In the zero-temperature Fermi-surface formulation implemented in \textsc{QuantumGeom}~\cite{Yu2021PRB}, the diagonal components of the conductivity tensor can be written as~\cite{ZimanBook,Madsen2006BoltzTraP}
\begin{equation}
\label{eq:sigma_diag_FS}
\frac{\sigma_{\alpha\alpha}}{\tau}=
\frac{e^{2}}{\hbar}
\sum_{n}
\int_{\mathrm{FS}_n}
\frac{dS}{(2\pi)^3}
\frac{v_{n,\alpha}^{2}(\mathbf{k})}
{|\mathbf{v}_{n}(\mathbf{k})|},
%\nonumber,
\qquad \alpha\in\{x,y,z\},
\end{equation}
where  
$\mathbf{v}_{n}(\mathbf{k} ) =
\frac{1}{\hbar}
\nabla_{\mathbf{k}}\varepsilon_{n\mathbf{k}}$ are band velocities, $\varepsilon_{n\bf{k}}$ are band energies, and the sum runs over all bands crossing the Fermi level $E_F$. Thus, $\sigma_{\alpha\alpha}/\tau$ is controlled not only by the density of states at the Fermi level, but also by the directional velocity weight carried by each Fermi-surface element.

\section{Results and Discussion}\label{results}

\subsection{Magnetic configurations and electronic fingerprints}\label{subsec:mag}

FeSb$_2$ crystallizes in the marcasite structure with nonmagnetic space group {\it Pnnm}. 
The structure, which consists of a mix of corner-sharing and edge-sharing FeSb$_6$ octahedra, has an orthorhombic primitive cell containing two formula units. 
We considered three collinear magnetically compensated states of FeSb$_2$: the AM state and two conventional antiferromagnetic states, AFM1 and AFM2, shown in Fig.~\ref{fig1}(a). 
Although all three configurations have zero net magnetization, their magnetic symmetries impose different constraints on the Bloch-band degeneracies. 
In the AM configuration, the magnetic order does not increase the size of the unit cell. 
Rather, the two Fe sites in the primitive cell have opposite spin. 
The AFM1 configuration differs from the AM configuration in that the cell size is doubled along [001], with alternating spin-up and spin-down Fe sites along the $c$ direction. 
In the AFM2 configuration, the cell is doubled along both [001] and [010]. 
Although AFM1 can be represented in a smaller supercell than AFM2, calculations employed the same $1\times 2\times 2$ supercell for both AFM structures to facilitate comparisons. 
In the AM state, the symmetry relation between opposite-spin sublattices maps $\mathbf{k}$ to a symmetry-related rotated momentum point rather than leaving $\mathbf{k}$ invariant. 
As a result, no antiunitary symmetry preserving the same $\mathbf{k}$ point enforces double degeneracy at generic $\mathbf{k}$, and momentum-dependent spin splitting is allowed despite the absence of a net moment. 
This mechanism is qualitatively distinct from ferromagnetic exchange splitting, which is tied to a nonzero net magnetization. 
In the AM state, the splitting already appears at the nonrelativistic exchange level and is therefore not set by the much smaller SOC scale~\cite{PhysRevX.12.031042,PhysRevX.12.040501,PhysRevLett.126.127701}.
Figure~\ref{fig1}(b) shows the orthorhombic Brillouin zone with high-symmetry points labeled.
The spin-polarized Fermi-surface map in Fig.~\ref{fig1}(c) exhibits alternating spin polarization across symmetry-related regions of the Brillouin zone. 
Correspondingly, the spin-resolved bands in Fig.~\ref{fig1}(d) show a clear lifting of the spin degeneracy along the $\Gamma$--$R'$--$R$--$\Gamma$ path. 
The coexistence of zero net magnetization and spin splitting at generic $\mathbf{k}$ in the no-SOC limit is a characteristic spectral hallmark of altermagnetism and parallels recent experimental observations in $\alpha$-MnTe and related systems~\cite{Krempaský2024,PhysRevLett.132.036702}. 
Taken together, Figs.~\ref{fig1}(c) and \ref{fig1}(d) show that the AM configuration of FeSb$_2$ displays the expected momentum-space signatures of altermagnetism, consistent with earlier theoretical proposals for FeSb$_2$ and recent high-field torque magnetometry and first-principles studies of its Fermi-surface electronic structure~\cite{Mazin2021,PhysRevB.111.075141}.
In contrast, both the AFM1 and AFM2 magnetic structures are bipartite lattices that are invariant under combined $\mathcal{T}\tau$ symmetry, where $\tau$ is a half-lattice translation within the supercell. 
This protects the Kramers spin degeneracy at each point in the Brillouin zone. 
Indeed, Figs.~\ref{fig1}(e,f) show spin-degenerate bands in
AFM1 and AFM2, even at {\bf k} points where the AM bands are spin split, consistent with the behavior expected for conventional compensated antiferromagnetic states~\cite{RevModPhys.90.015005}. 
The contrast between Fig.~\ref{fig1}(d) and Figs.~\ref{fig1}(e,f) therefore highlights the symmetry distinction between the AM state and conventional compensated antiferromagnetic order in FeSb$_2$. 
Related unconventional magnetic configuration and symmetry-allowed spin splitting in FeSb$_2$ were also investigated previously in the context of alloy-stabilized altermagnetism~\cite{Mazin2021}.

\iffalse
\begin{table}[b]
\caption{Relative total energies of FeSb$_2$ under triaxial tensile strain. For each strain value, the lowest-energy state among the configurations considered is set to zero. Energies are reported in meV per Fe$_2$Sb$_4$ primitive cell. }
\label{tab1}
\centering
\renewcommand{\arraystretch}{1.25}
\setlength{\tabcolsep}{8pt}
\begin{tabular}{c|ccccc}
\hline\hline
$\varepsilon$ (\%) & AM & AFM1 & AFM2 & NM & FM \\
\hline
0 & 17.0 & \textbf{0.0} & 0.3 & 4.0 & 4.0 \\
1 & \textbf{0.0} & 27.0 & 26.9 & 45.0 & 16.0 \\
3 & \textbf{0.0} & 93.8 & 70.8 & 182.0 & 19.0 \\
5 & \textbf{0.0} & 114.0 & 109.1 & 346.0 & 38.0 \\
\hline\hline
\end{tabular}
\end{table}
\fi

\subsection{Effect of lattice strain}\label{subsec:strain}

Having established the symmetry fingerprints of the candidate magnetic states, we now turn to their relative energetics. 
For the optimized structures at zero pressure, the total energies of nonmagnetic (NM), ferromagnetic (FM), AM, AFM1 and AFM2 FeSb$_2$ are all calculated to lie within a range of about 20 meV per primitive cell, where the primitive cell contains 2 Fe atoms. 
The AFM2 structure is lowest in energy, followed by AFM1, AM, NM, and FM. 
An energy landscape with multiple local minima that are close in energy suggests the possibility of tuning the magnetic ground state with small perturbations. 
For example, alloying has been predicted to provide a route to stabilizing AM FeSb$_2$~\cite{Mazin2021}. 
Here we investigate the effect of lattice strain.

\begin{figure}[tb]
    \centering
    \includegraphics[width=\linewidth]{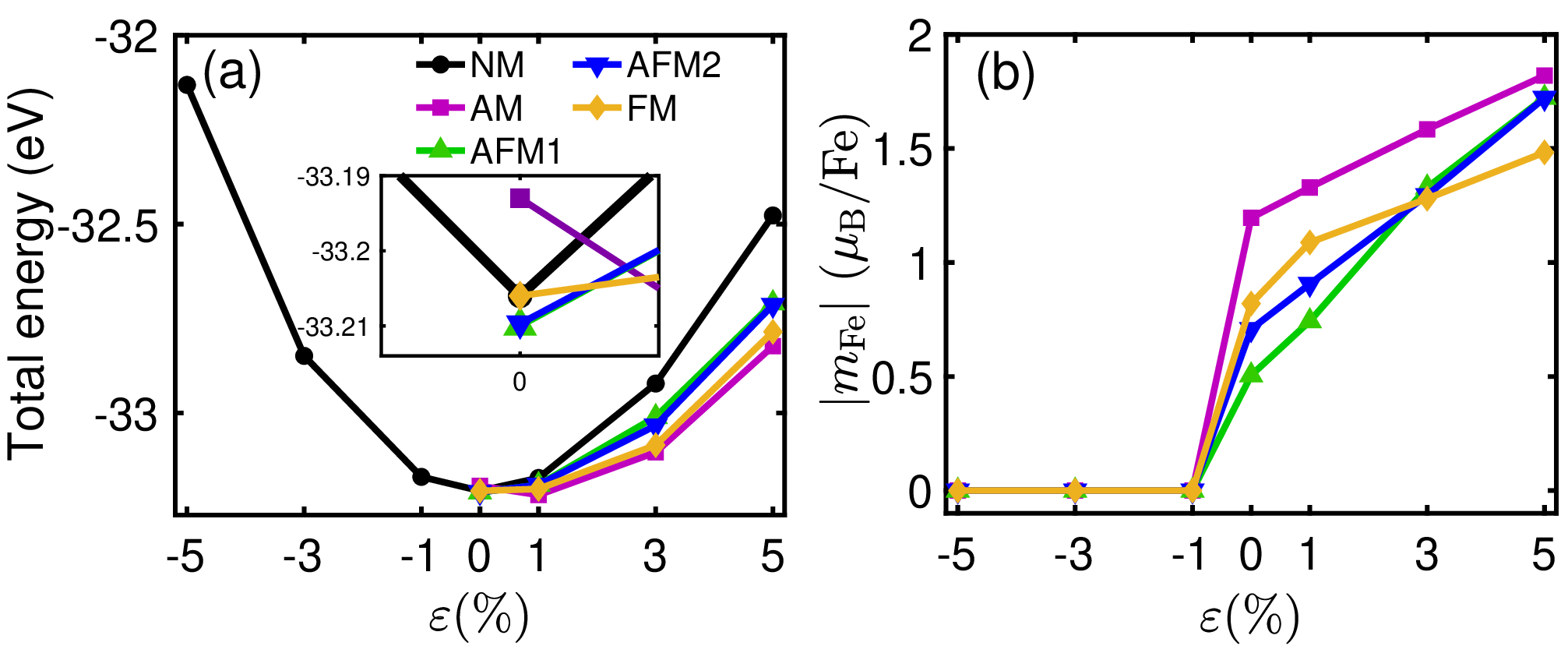}
\caption{(a) Total energies of the NM, AM, AFM1, AFM2, and FM states of FeSb$_2$ as a function the volumetric strain parameter $\varepsilon$. The energies refer to a Fe$_2$Sb$_4$ unit cell.
The inset enlarges the region around $\varepsilon=0$  
to highlight the near degeneracies and strain-induced energy reordering. (b) Corresponding local magnetic moments, shown as the average absolute Fe-site moment $\langle |m_{\mathrm{Fe}}| \rangle$ in units of $\mu_B$/Fe. 
The total moment of the compensated AM, AFM1, and AFM2 states remains zero. 
}
    \label{fig_totEnergy}
\end{figure}

\begin{figure*}[htb]
    \centering
\includegraphics[width=0.95\linewidth]{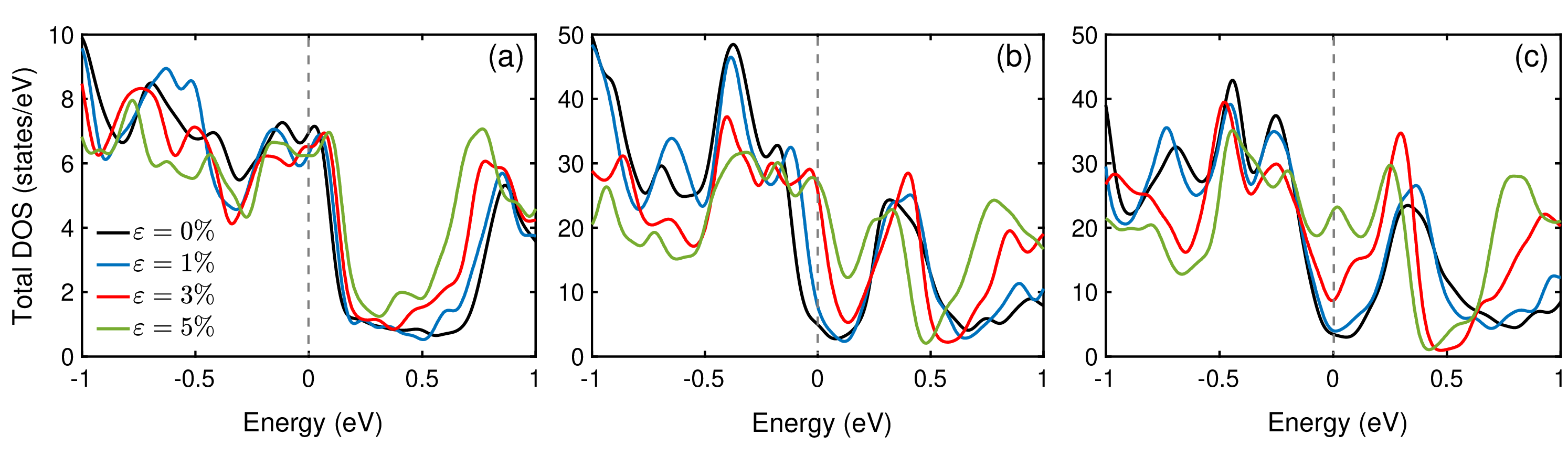}
\caption{Strain evolution of the total density of states in FeSb$_2$.
Total density of states (TDOS) of FeSb$_2$ in the (a) AM, (b) AFM1, and (c) AFM2 configurations for volumetric tensile strain with $\varepsilon=0\%, 1\%, 3\%$, and $5\%$. 
For each phase and $\varepsilon$ value, the energy is referenced to the corresponding self-consistent Fermi level, marked by the vertical dashed line at $E_F=0$.}
\label{fig_dos}
\end{figure*}

The optimized lattice parameters for NM FeSb$_2$, $a=5.80$~\AA, $b=6.51$~\AA, and $c=3.19$~\AA, were used to define a reference volume, $V_0^{NM}=120.66$~\AA$^3$. 
To explore the effects of volumetric strain, we first constructed a uniformly expanded (or compressed) orthorhombic cell by scaling all three lattice constants by the same factor $(1+\varepsilon)$, 
yielding a target volume $V(\varepsilon)=(1+\varepsilon)^3V_0^{NM}$.
The target volume was then held fixed while the cell shape and all internal atomic coordinates were relaxed with symmetry constraints disabled. 
Because the same reference volume $V_0^{NM}$ was used to define $V(\varepsilon)$ for each magnetic state, and because the cell shape was optimized while keeping the volume constant, $\varepsilon$ labels the volume rather than the actual linear strain along each crystalline axis. 
Figure~\ref{fig_totEnergy}(a) shows the total energies (per Fe$_2$Sb$_4$) of the five magnetic phases as a function of $\varepsilon$.
At $\varepsilon=0$, corresponding to $V=V_0^{NM}$, AFM1 and AFM2 are nearly degenerate and form the lowest-energy compensated states among the configurations considered. 
FM and NM states lie only about $4$~meV higher, while the AM solution is higher by about $17$~meV.
A modest volume expansion corresponding to $\varepsilon = 0.01$ reverses this ordering and makes AM the lowest-energy solution among the tested magnetic configurations. 
With further tensile strain, AM remains lowest in energy, FM is the closest competing phase, and the NM solution is progressively destabilized. 
Figure~\ref{fig_totEnergy}(b) shows the corresponding local magnetic moments, where the plotted quantity is 
the average absolute value of the local moment on the Fe sites, $\langle |m_{\mathrm{Fe}}| \rangle$, reported in units of $\mu_B$/Fe. 
Although the local moments vary with strain, the AM, AFM1, and AFM2 states remain magnetically compensated: their total magnetizations vanish due to cancellation between opposite Fe sublattices.
The increase of $\langle |m_{\mathrm{Fe}}| \rangle$ with tensile strain therefore reflects a strengthening of the local exchange polarization, not the development of a net magnetization. 
On the other hand, compressive strain quenches the local moments. 
At volumes with $\varepsilon \lesssim -0.01$, all of the tested initial magnetic configurations relax to the NM state, as reflected by the collapse of $\langle |m_{\mathrm{Fe}}| \rangle$ to zero. 
For this reason, and because our main interest is the strain-induced stabilization of altermagnetism, the remainder of this paper focuses on the tensile-strain regime. 
The transport analysis is applied to the compensated magnetic states in order to compare the altermagnetic band-transport response to that of the conventional AFM references. 
The FM phase, although energetically competitive at finite tensile strain, is not used as a transport reference because it carries a finite net moment and hence is easily distinguishable from the magnetically compensated phases.
Figure~\ref{fig_dos} shows the effect of tensile strain on the total density of states (TDOS) of FeSb$_2$ for the AM, AFM1, and AFM2 configurations. 
Within the present DFT treatment, all three phases at all tensile strains considered have finite TDOS at the Fermi level, $N(E_F)$, indicating metallic behavior. 
Nevertheless, there are differences, even at
the reference volume ($\varepsilon = 0$).
In particular, at $\varepsilon = 0$, the TDOS at the Fermi level in the AM phase is large, while in the AFM1 and AFM2 phases, the Fermi level lies within a deep minimum in the TDOS.
The position of the Fermi level in a pseudogap lowers the one-electron energy in the AFM phases, contributing to their relative stability compared to the AM phase~\cite{Mazin2021}. 

\begin{figure*}[htb]
    \centering
    \includegraphics[width=0.45\linewidth]{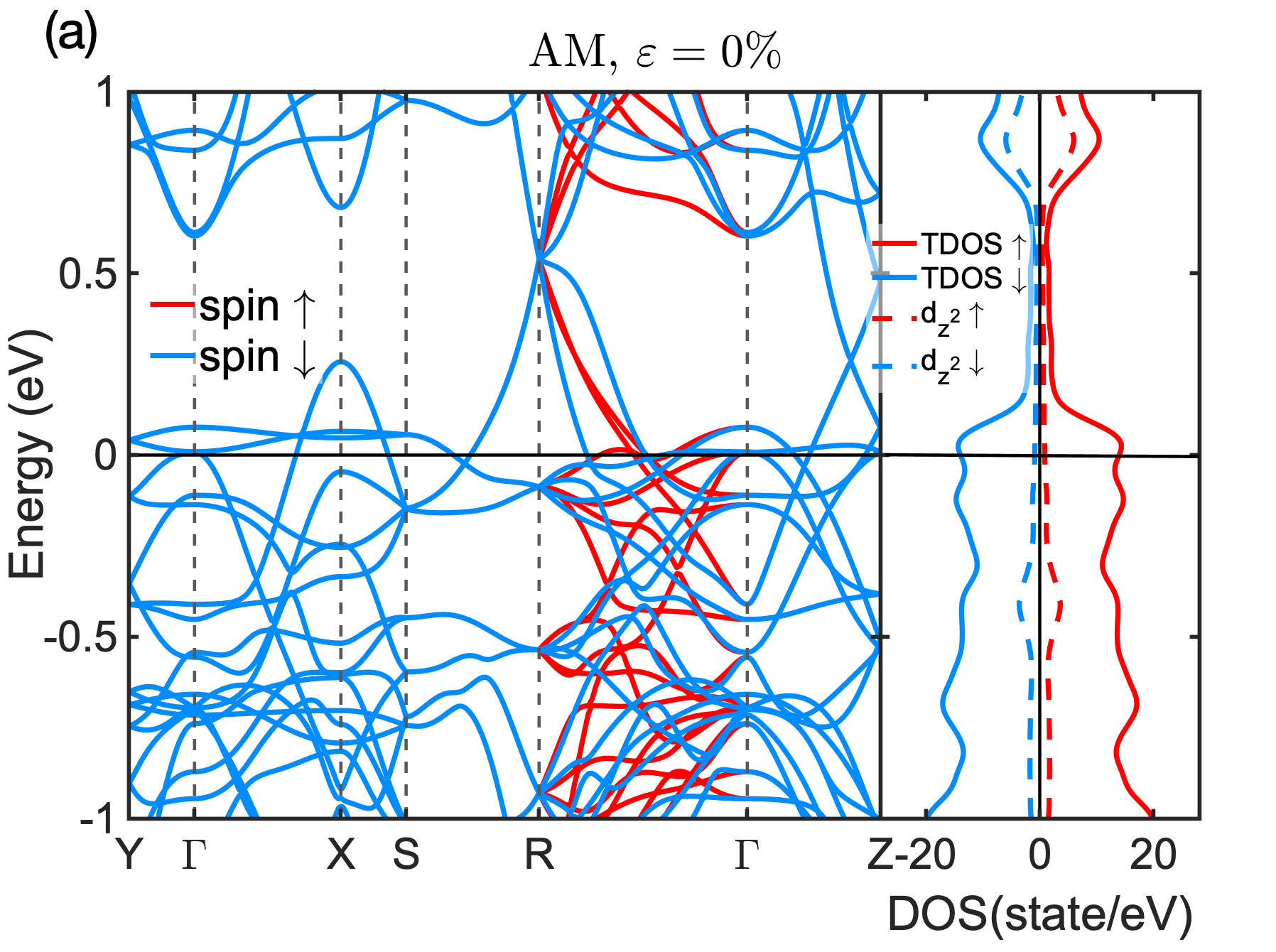}
    \includegraphics[width=0.45\linewidth]{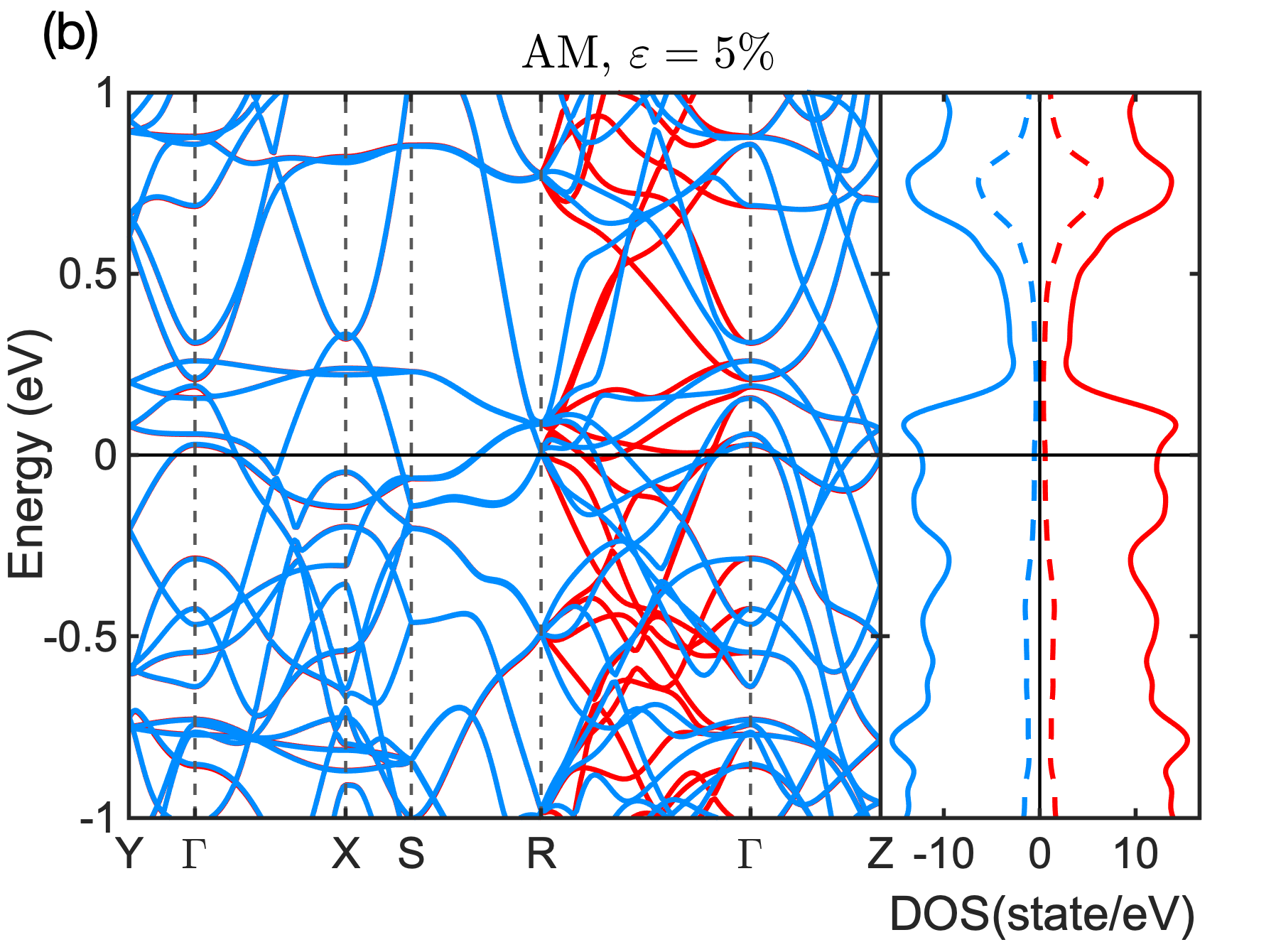}\\
    \includegraphics[width=0.45\linewidth]{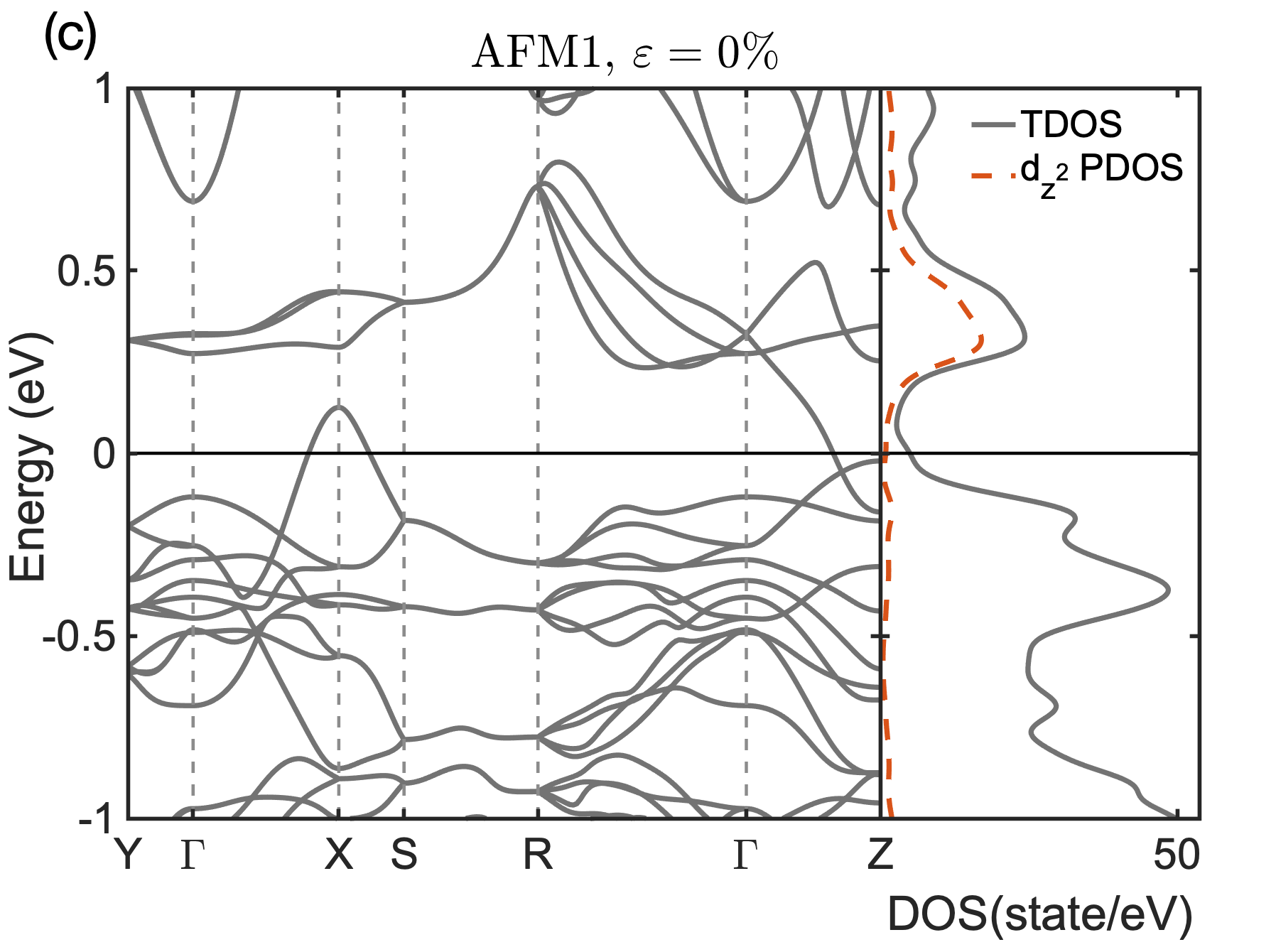}
    \includegraphics[width=0.45\linewidth]{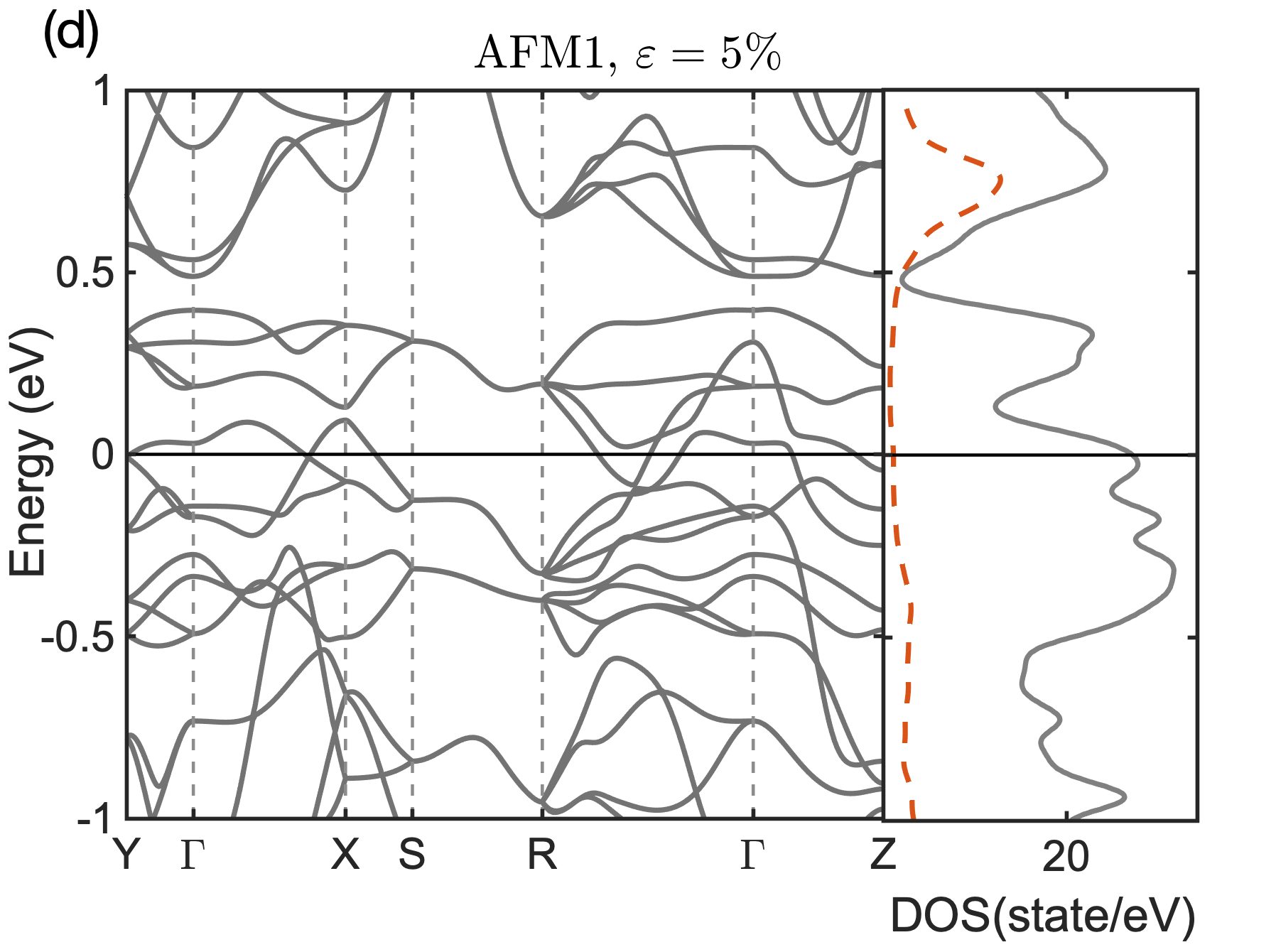}
\caption{Strain evolution of the band structure of AM and AFM1 FeSb$_2$. Panels (a,b) show AM at $\varepsilon=0\%$ and $5\%$, and panels (c,d) show AFM1 at the same strains. To facilitate comparison, the AM and AFM1 bands are both plotted in a folded Brillouin zone corresponding to a $1\times2\times2$ supercell.
 Red and blue bands denote opposite AM spin projections, while AFM1 bands are spin-degenerate in the absence of SOC. The Fermi level is set to 0.  The bands are shown together with the corresponding TDOS (solid) and Fe-$d_{z^2}$ projected DOS (dashed). In the AM panels, the  DOS for the two spin projections are plotted with opposite sign. }
    \label{fig:am_bands_strain}
\end{figure*}

To understand the pseudogap and how it depends on magnetic structure and strain, it is helpful to look at the band structure and orbital character near the Fermi energy, as shown in Fig.~\ref{fig:am_bands_strain}. 
In this figure, the AM bands are plotted in the same $1\times2\times2$ supercell representation as AFM1; this folds the primitive-cell AM bands into the supercell Brillouin zone without changing the compensated altermagnetic order. 
For AFM1, the peak in the TDOS around 0.3 eV above $E_F$ arises from narrow $d_{z^2}$ bands in a pseudogap between a complex of occupied $d$ bands below $E_F$ and a complex of unoccupied $d$ bands starting about 0.7 eV above $E_F$. 
The narrowness of the $d_{z^2}$ bands (and other $d$ bands as well) is due to the pattern of alternating spins on neighboring Fe sites in the AFM configuration, which reduces orbital overlap within each spin channel.
In the AM configuration, on the other hand, Fe sites form same-spin chains along the $c$ direction, which broadens and splits the $d_{z^2}$ bands. 
Since the bonding $d_{z^2}$ bands become occupied, the Fermi level shifts below the edge of the pseudogap.
When the volume is expanded (up to $\varepsilon = 0.05$), the pseudogap narrows as the splitting between $d$ bands decreases. 
In the AM phase, the Fermi level remains below the pseudogap as the lattice is expanded, while in the AFM1 phase, the $d_{z^2}$ bands start to overlap with the surrounding bands, filling in the pseudogap and increasing the TDOS at the Fermi level. 
A similar picture holds for the strain response of AFM2. 
In both AFM solutions, tensile strain generally enhances the near-$E_F$ spectral weight, increasing $N(E_F)$ and reducing their energetic favorability relative to the AM phase.
Figure~\ref{fig:am_bands_strain} also shows that the strained structures retain their magnetic symmetry.
Although individual band positions shift with strain and the low-energy dispersions are reorganized, the AM solution retains spin-split bands at generic {\bf k}-points while remaining fully compensated, and the conventional AFM solutions maintain doubly degenerate bands. 
Together with the energetic trends  in Fig.~\ref{fig_totEnergy}, these results show that volumetric tensile strain favors the AM solution from $\varepsilon=0.01$ onward while preserving its hallmark altermagnetic electronic structure. 
The strain induces a transition between two metastable magnetic solutions rather than a transition driven by a change in crystal symmetry~\cite{PhysRevB.109.144421}.

\begin{figure}[t]
    \centering
\includegraphics[width=\linewidth]{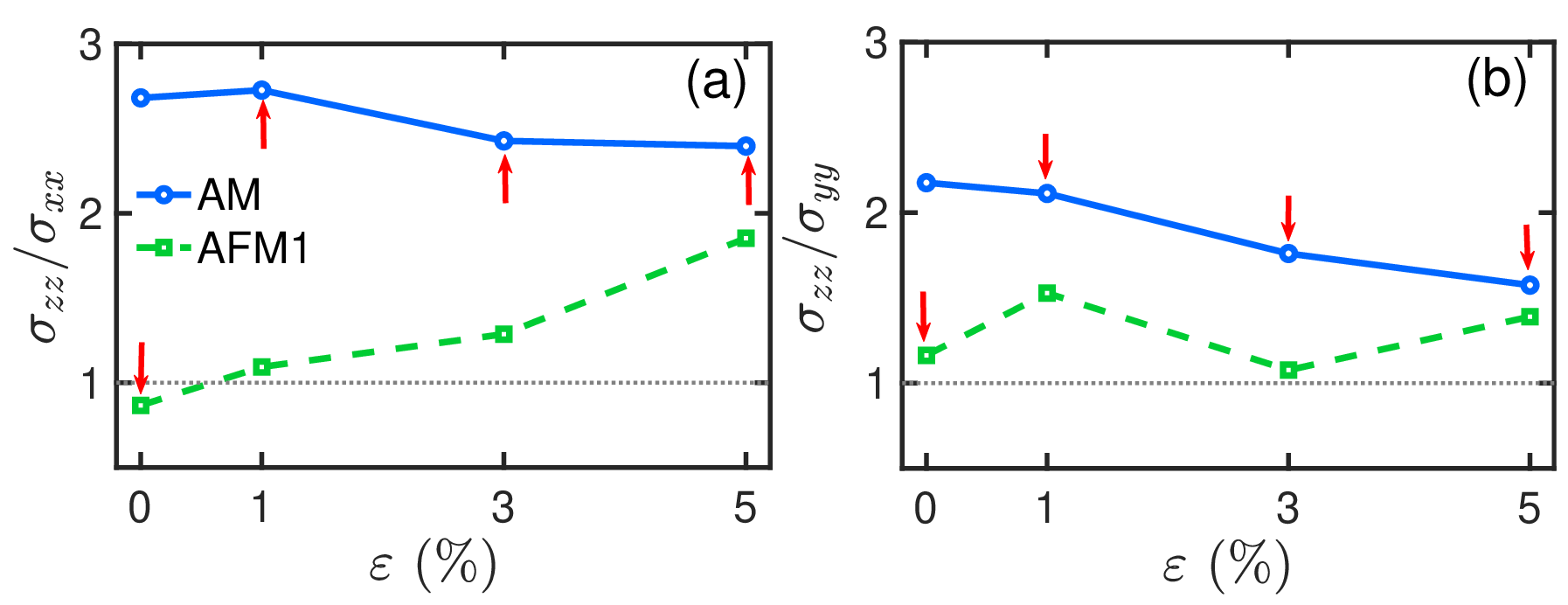}
\caption{Evolution of the longitudinal conductivity anisotropy in FeSb$_2$ under volumetric tensile strain.
Panels (a) and (b) show the ratios $\sigma_{zz}/\sigma_{xx}$ and $\sigma_{zz}/\sigma_{yy}$, respectively, for the AM and AFM1 configurations  over the full volume-expansion series. The red arrows indicate the magnetic ground-state configuration at the corresponding strain.}
    \label{fig_sigma_CL}
\end{figure}

We next examine how magnetic order and strain affect the longitudinal transport anisotropy of FeSb$_2$. 
Figure~\ref{fig_sigma_CL} shows the ratios $\sigma_{zz}/\sigma_{xx}$ and $\sigma_{zz}/\sigma_{yy}$ for the AM and AFM1 configurations over the full volume-expansion series.
At $\varepsilon=0$, the AM phase exhibits dominant $z$-axis response with $\sigma_{zz}/\sigma_{yy}, \sigma_{zz}/\sigma_{xx} > 2$, while transport in the AFM1 phase is much more isotropic, with $\sigma_{zz}/\sigma_{yy}, \sigma_{zz}/\sigma_{xx} \sim 1$. 
This difference in anisotropy can be understood from the pattern of magnetic moments in real space, as discussed above.
In the AM structure, nearest-neighbor Fe sites along the orthorhombic $c$ direction carry parallel moments, with a separation of about $3.2$~\AA\ in the reference structure. 
By contrast, same-spin Fe sites along the $a$ and $b$ directions are not connected directly,
but are instead reached through longer paths involving opposite-moment Fe sites. 
Without SOC, the spin channels do not mix, so the AM structure provides a more direct same-spin Fe-$d$ hopping network along $c$. 
This naturally favors stronger $k_z$ dispersion and larger $v_z^2$ weight in the conductivity integral in Eq.~\eqref{eq:sigma_diag_FS}.
In AFM1 and AFM2, this same-spin Fe chain along $c$ is interrupted, consistent with the absence of a robust $\sigma_{zz}$-dominated response in the conventional AFM references. 
In reciprocal space, the difference in anisotropy manifests in the Fermi surfaces of the AM and AFM phases, as shown in Appendix~\ref{app:FS}.  
Sections of the AM Fermi surface form relatively flat sheets normal to $k_z$, consistent with large $v_z$ contributions to the conductivity. 
In contrast, the AFM1 Fermi surface consists of small pockets that are more isotropic. 
Under tensile strain, the conductivity anisotropy of the AM configuration remains robust. 
At all strains considered, both $\sigma_{zz}/\sigma_{xx}$ and $\sigma_{zz}/\sigma_{yy}$ remain well above unity, establishing dominant $z$-axis response. 
This is consistent with the relatively modest influence of strain on the AM bands near $E_F$ (Fig.~\ref{fig:am_bands_strain}).
At low tensile strains, longitudinal transport in the AFM1 remains roughly isotropic. 
At $\varepsilon=0.05$, $\sigma_{zz}$ starts to dominate, which reflects the strain-induced reorganization of near-$E_F$ states seen in Fig.~\ref{fig:am_bands_strain}. 
The auxiliary AFM2 data in Appendix~\ref{app:afm2_lowstrain} show that, similar to AFM1, at strains near or below the magnetic crossover, 
the conductivity remains relatively isotropic. 
Therefore the pronounced $z$-axis transport dominance of the AM phase is not a generic consequence of magnetic compensation or orthorhombic lattice anisotropy alone, but is tied to the velocity-weighted near-Fermi electronic structure of the altermagnetic state, which differs significantly from that of the conventional AFM states due to the pattern of magnetic moments in real space.

\subsection{Spin--orbit-coupling effects}
\label{subsec:soc}
To assess the impact of SOC on the longitudinal transport anisotropy, we compare results with and without SOC for the AM and AFM1 states at $\varepsilon=0.01$ volumetric tensile strain, near the magnetic crossover.
The conductivity prefactors are listed in Table~\ref{tab:soc_transport}. 
In the AM state, the dominant $z$-axis response remains even when SOC is included,
consistent with the altermagnetic spin splitting being driven by nonrelativistic exchange and magnetic crystal symmetry rather than by SOC~\cite{PhysRevX.12.031042,PhysRevLett.126.127701}.
Including SOC changes the absolute prefactors and reduces the anisotropy ratios modestly,  %
but preserves the same AM hierarchy
$\sigma_{zz}>\sigma_{yy} > \sigma_{xx}$.
AFM1 behaves differently, following $\sigma_{zz}>\sigma_{xx}>\sigma_{yy}$ both with and without SOC, while the ratio $\sigma_{zz}/\sigma_{xx}$ is much closer to unity than in the AM phase. 
Thus, within the CRTA and for moments initialized along [010], SOC 
does not change the qualitative transport distinction between the AM and conventional AFM reference.

\begin{table}[t]
\caption{Effect of SOC on the diagonal longitudinal conductivity prefactors in AM and AFM1 FeSb$_2$ at $\varepsilon=1\%$ volumetric tensile strain.
 Conductivity prefactors are reported in units of $10^{15}\,\Omega^{-1}\mathrm{cm}^{-1}\mathrm{s}^{-1}$.
 The no-SOC values are spin-summed, while the SOC calculations were initialized with magnetic moments along the orthorhombic $b$ direction.}
\label{tab:soc_transport}
\centering
\renewcommand{\arraystretch}{1.15}
\begin{tabular}{l l c c c c c}
\hline\hline
State & calculation & $\sigma_{xx}/\tau$ & $\sigma_{yy}/\tau$ & $\sigma_{zz}/\tau$
& $\sigma_{zz}/\sigma_{xx}$ & $\sigma_{zz}/\sigma_{yy}$ \\
\hline
AM & no SOC & 539 & 697 & 1474 & 2.73 & 2.11 \\
AM & SOC    & 634 & 782 & 1373 & 2.17 & 1.76 \\
\hline
AFM1 & no SOC & 240 & 172 & 263 & 1.09 & 1.53 \\
AFM1 & SOC    & 231 & 158 & 281 & 1.22 & 1.78 \\
\hline\hline
\end{tabular}
\end{table} 
\section{Conclusion}\label{conclusions}

\begin{figure*}[t]
    \centering
    \includegraphics[width=\linewidth]{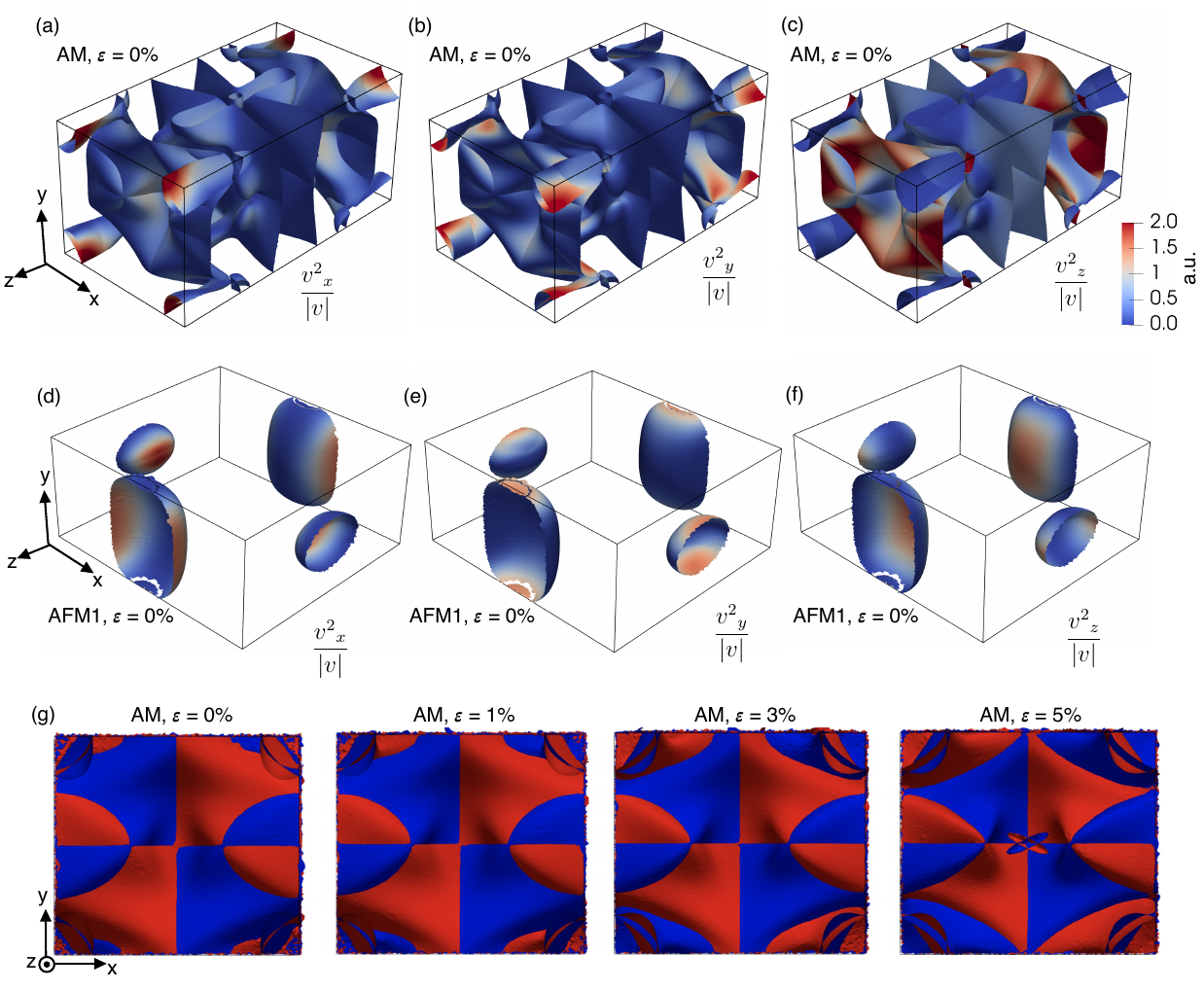}
 \caption{Strain evolution of the AM spin-polarized Fermi surface.
(a)--(c) Velocity-weighted Fermi-surface maps of the unstrained AM configuration, colored by $v_x^2/|\mathbf{v}|$, $v_y^2/|\mathbf{v}|$, and $v_z^2/|\mathbf{v}|$, respectively. 
(d)--(f) Corresponding maps for the unstrained AFM1 configuration. 
The plotted scalar $v_\alpha^2/|\mathbf{v}|$ visualizes the directional velocity weight entering the Fermi-surface expression for $\sigma_{\alpha\alpha}/\tau$ in Eq.~\eqref{eq:sigma_diag_FS}.
(g) Spin-projected Fermi-surface maps of the AM configuration for $\varepsilon=0\%, 1\%, 3\%,$ and $5\%$ under the volumetric-expansion protocol. Red and blue denote opposite spin projections along the chosen collinear spin quantization axis. The alternating momentum-space spin-polarization pattern persists across the expansion series, showing that the AM spin texture survives in the strain-stabilized regime.}
    \label{fig:am_fs_spin_strain}
\end{figure*}
\appendix

This work shows that volumetric tensile strain can reorganize the competition among compensated magnetic states in FeSb$_2$ and select an altermagnetic solution with a distinct longitudinal transport response. 
At the reference volume, the conventional compensated AFM states are favored among the configurations considered, while the AM state is metastable. 
Under tensile strain, however, this ordering reverses and the AM solution becomes energetically favored.
The strain-stabilized AM phase retains its characteristic no-SOC altermagnetic electronic structure, including momentum-dependent spin splitting and metallic near-Fermi bands. 
Its longitudinal conductivity prefactors show a robust $z$-axis dominance
throughout the tensile strain series. 
This strong uniaxial anisotropy 
is not observed in the conventional AFM configurations near or below the magnetic crossover,
indicating that the dominant $z$-axis response is tied to the real-space pattern of magnetic moments in
the AM phase rather than to magnetic compensation or orthorhombic lattice anisotropy alone.
Representative calculations including the spin-orbit interaction show that SOC renormalizes the conductivity prefactors and anisotropy ratios but preserves the dominant $z$-axis response that is present in the AM electronic structure without SOC. 
Overall, these results identify volumetric tensile strain as a route to stabilize the AM configuration in FeSb$_2$ and suggest that changes in the conductivity anisotropy could serve as an experimental indicator of the magnetic transition. 

\section*{Acknowledgments}
This work was supported by the U.S. National Science Foundation under Grant No.~NSF-DMR-2440337 and NSF-DMR-2349397. The calculations were performed on Bridges-2 at the Pittsburgh Supercomputing Center through allocations PHY250051p and MAT230004p from the Advanced Cyberinfrastructure Coordination Ecosystem Services \& Support (ACCESS) program, which is supported by National Science Foundation (US) grants \#2138259, \#2138286, \#2138307, \#2137603, and \#2138296.

\section*{Data Availability}
The data that support the findings of this article are available from the corresponding author upon reasonable request.

\section{Additional Fermi-surface and conductivity-prefactor data}\label{app:FS}

The differences in conductivity anisotropy between the AM and AFM phases is related to differences in the geometry of the Fermi surfaces. 
Panels (a)--(f) of Figure~\ref{fig:am_fs_spin_strain} show velocity-weighted Fermi-surface maps for the  AM and AFM1 configurations at the reference volume ($\epsilon=0$).
The color indicates $v_\alpha^2/|\mathbf{v}|$, the directional velocity factor entering the longitudinal conductivity integral.
For the unstrained AM state, Figs.~\ref{fig:am_fs_spin_strain}(a)--(c) provide a qualitative visualization of the dominant $z$-axis velocity weight.
Compared to the velocity maps for the other directions, the $v_z^2/|\mathbf{v}|$ map shows larger values of the velocity weight maintained on larger regions of the Fermi surface. 
By contrast, the AFM1 maps in Figs.~\ref{fig:am_fs_spin_strain}(d)--(f) show similar maximum values of velocity weights covering similar-sized regions of the Fermi surface, consistent with the relatively isotropic conductivity calculated for the AFM1 state.
Figure~\ref{fig:am_fs_spin_strain}(g) shows the strain evolution of the spin-projected AM Fermi surface. 
Even as new Fermi-surface sheets emerge with tensile strain,
the alternating red-blue texture persists, showing that the strain-stabilized AM state retains its scalar-relativistic exchange-split Fermi surface. 

\begin{figure}[t]
    \centering
\includegraphics[width=\linewidth]{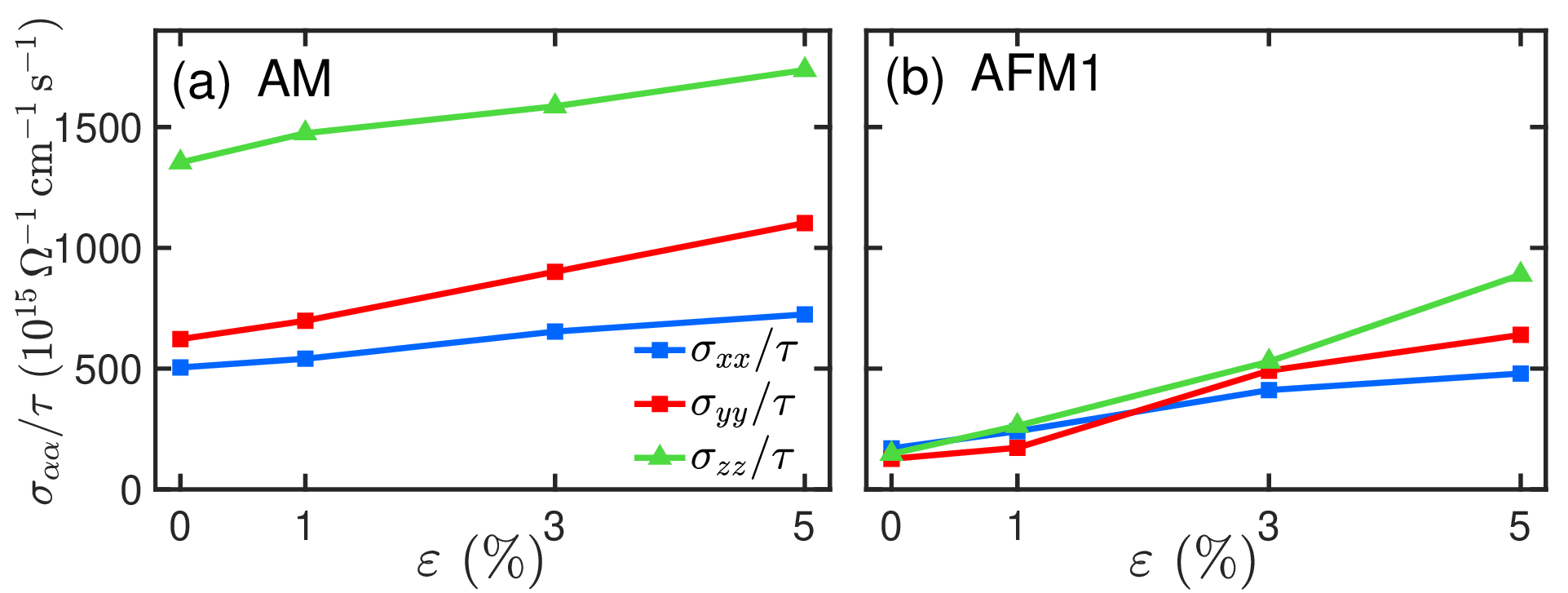}
  \caption{Strain dependence of the diagonal longitudinal conductivity prefactors of FeSb$_2$.
The quantities $\sigma_{xx}/\tau$, $\sigma_{yy}/\tau$, and $\sigma_{zz}/\tau$ are shown for (a) the AM configuration and (b) the AFM1 configuration over the full volume-expansion series. }

    \label{fig_sigma_ii}
\end{figure}

To understand the evolution of the anisotropic conductivity modulated by the strain, we examine the diagonal conductivity prefactors $\sigma_{\alpha\alpha}/\tau$ underlying the anisotropy ratios in Fig.~\ref{fig_sigma_CL}. 
In the AM configuration, all three conductivity prefactors increase with expansion, but the $z$ component remains well separated from the transverse components throughout the strain series, as shown in Fig.~\ref{fig_sigma_ii}(a). 
The strain-induced reduction of the AM anisotropy ratios in Fig.~\ref{fig_sigma_CL} therefore does not come from a loss of the $z$-axis channel; rather, it reflects the faster relative growth of the $x$ and $y$ components as the lattice expands.
AFM1 shows a different component-level trend. 
Its conductivity prefactors also increase with strain, but the three directions remain much closer in magnitude than in the AM case, and their relative evolution is less systematic. 
In particular, AFM1 does not maintain the ordered separation $\sigma_{zz}/\tau>\sigma_{yy}/\tau>\sigma_{xx}/\tau$ seen in AM. 
Overall, Fig.~\ref{fig_sigma_ii} demonstrates that the AM anisotropy originates from a robust directional separation of conductivity components, whereas AFM1 and AFM2 (see Appendix~\ref{app:afm2_lowstrain}) show weaker and less systematic conductivity anisotropy.
Since the ground-state transition from AFM1 to AM occurs around $\varepsilon=0.01$, we expect a drastic change in the conductivity anisotropy when the transition occurs. 

\section{Low-strain AFM2 transport}
\label{app:afm2_lowstrain}
AFM2 is retained in the main text as a secondary conventional AFM reference in the magnetic-energetics and band-symmetry analysis. 
Here we report its conductivity prefactors at $\varepsilon=0$ and 0.01, spanning the crossover between the conventional-AFM regime and the strain-stabilized AM regime. 
Table~\ref{tab:afm2_lowstrain_transport} lists the diagonal conductivity prefactors $\sigma_{\alpha\alpha}/\tau$ and the corresponding anisotropy ratios calculated within the same constant-relaxation-time framework used throughout the main text. 
In contrast to the AM phase, the AFM2 conductivity components along the three crystallographic directions exhibit relatively weak anisotropy, and in particular,  $\sigma_{zz} \lesssim \sigma_{xx}, \sigma_{yy}$ near and below the magnetic crossover. 
Together with the AFM1 comparison in the main text, these results indicate that the pronounced $z$-axis transport dominance in the AM phase is not a generic consequence of magnetic compensation, orthorhombic lattice anisotropy, or volumetric tensile strain alone. 
Instead, it originates from the AM magnetic structure, which leads to differences in its velocity-weighted near-Fermi electronic structure compared to that of the conventional AFM configurations.

\begin{table}[t]
\caption{Low-strain AFM2 conductivity prefactors and anisotropy ratios. The conductivity prefactors,  $\sigma_{\alpha\alpha}/\tau$, are reported in units of $10^{15}\,\Omega^{-1}\mathrm{cm}^{-1}\mathrm{s}^{-1}$.  AFM2 is used here as a low-strain conventional-AFM reference near the magnetic crossover.}
\label{tab:afm2_lowstrain_transport}
\centering
\renewcommand{\arraystretch}{1.15}
\setlength{\tabcolsep}{6pt}
\begin{tabular}{c c c c c c}
\hline\hline
$\varepsilon$ (\%) & $\sigma_{xx}/\tau$ & $\sigma_{yy}/\tau$ & $\sigma_{zz}/\tau$
& $\sigma_{zz}/\sigma_{xx}$ & $\sigma_{zz}/\sigma_{yy}$ \\
\hline
0 & 86 & 106 & 83 & 0.97 & 0.78 \\
1 & 83 & 71  & 60 & 0.73 & 0.85 \\
\hline\hline
\end{tabular}
\end{table}

\section{Effects of uniaxial strain}
\label{app:uniaxial}

Besides the volumetric tensile strain discussed in the main text, we also examined the transport response under uniaxial strain. 
The strain was applied separately along the orthorhombic $a$, $b$, and $c$ axes at $\varepsilon=0.01, 0.03,$ and 0.05, starting from the lattice constants for the relaxed nonmagnetic reference structure. 
For the strain along a given axis, that lattice constant was increased by $(1+\varepsilon)$, while the other two lattice constants were kept fixed at their unstrained reference values. 
The strained lattice vectors were held fixed during relaxation, and only the internal atomic coordinates were relaxed, with symmetry constraints disabled and the same force threshold used in the volumetric-strain calculations. 
Because different uniaxial strain directions and magnitudes contain different elastic contributions, the magnetic configurations are compared only within the same fixed strain condition. 
In Table~\ref{tab:uniaxial_energy}, the energy of the lowest-energy state among AM, AFM1, and AFM2 is set to zero separately for each row, and the other entries are reported as relative energies normalized to the Fe$_2$Sb$_4$ reference cell. 
\begin{table}[htb]
\caption{Relative total energies of compensated magnetic states under uniaxial tensile strain. For each strain condition, the lowest-energy state among the AM, AFM1, and AFM2 configurations is set to $\Delta E=0$. Energies are reported in meV per Fe$_2$Sb$_4$ reference cell.}
\label{tab:uniaxial_energy}
\centering
\renewcommand{\arraystretch}{1.12}
\setlength{\tabcolsep}{7pt}
\begin{tabular}{lccc}
\hline\hline
Strain condition & AM & AFM1 & AFM2 \\
\hline
$1\%$ along $a$ & 18.6 & 10.7 & 0.0 \\
$1\%$ along $b$ & 16.0 & 10.7 & 0.0 \\
$1\%$ along $c$ & 15.4 & 13.7 & 0.0 \\
\hline
$3\%$ along $a$ & 47.2 & 15.3 & 0.0 \\
$3\%$ along $b$ & 126.0 & 14.1 & 0.0 \\
$3\%$ along $c$ & 21.7 & 22.3 & 0.0 \\
\hline
$5\%$ along $a$ & 0.0 & 41.1 & 20.8 \\
$5\%$ along $b$ & 0.0 & 60.3 & 42.6 \\
$5\%$ along $c$ & 0.0 & 64.3 & 40.6 \\
\hline\hline
\end{tabular}
\end{table}
Compared with the tensile strain series in Fig.~\ref{fig_totEnergy}, where the AM solution already becomes favored at $\varepsilon=0.01$, the uniaxial data show a delayed stabilization of AM. 
At both $\varepsilon= 0.01$ and $0.03$ uniaxial strain, AFM2 remains the lowest-energy solution among the three magnetically compensated configurations considered, regardless of the orthorhombic crystallographic axis along which the strain is applied. 
This behavior reflects the fact that for the same value of $\varepsilon$, single-axis tension is a weaker and more anisotropic structural perturbation than the volumetric tensile strain.
The crossover to AM occurs between $\varepsilon=0.03$ and $0.05$ uniaxial expansion. When $\varepsilon = 0.05$, AM is the lowest-energy solution for uniaxial strain along any of the orthorhombic axes, with AFM2 remaining the nearest competing state in all three cases. 
Thus, stabilization of AM order does not require isotropic expansion, but can be achieved with uniaxial expansion as well.

For a representative transport comparison, we consider the $\varepsilon=0.05$ uniaxial deformation along the orthorhombic $a$ axis, where AM is already the lowest-energy solution among the three compensated magnetic configurations in Table~\ref{tab:uniaxial_energy}. 
The corresponding conductivity values are summarized in Table~\ref{tab:uniaxial_transport}.
The AM state shows a pronounced longitudinal anisotropy, with the spin-summed conductivity prefactors following $\sigma_{zz}>\sigma_{yy}>\sigma_{xx},$ and $(\sigma_{zz}/\sigma_{xx},\,\sigma_{zz}/\sigma_{yy})=(3.69,\,2.72)$. 
These ratios exceed the corresponding $\varepsilon=0.05$ volumetric tensile strain AM values, $(2.40,\,1.57)$, shown in Fig.~\ref{fig_sigma_CL}, indicating that uniaxial strain along $a$ further enhances the AM transport anisotropy. 
The conventional AFM configurations show a much weaker %directional separation 
conductivity anisotropy in the same geometry, with their three conductivity components remaining substantially closer to one another than in the AM state. 
In AFM2, for example, the $z$ component only slightly exceeds the $x$ component. This trend is consistent with the volumetric tensile strain results in the main text: in both strain protocols, the AM state exhibits a much stronger and more robust longitudinal transport anisotropy than the conventional AFM configurations.

\begin{table}[b]
\caption{Collinear no-SOC longitudinal conductivity prefactors for $5\%$ uniaxial strain along $a$.
The quantities $\sigma_{\alpha\alpha}/\tau$ are reported in units of $10^{15}\,\Omega^{-1}\mathrm{cm}^{-1}\mathrm{s}^{-1}$. }
%For AM, the listed value is the spin-summed charge conductivity.}
\label{tab:uniaxial_transport}
\centering
\renewcommand{\arraystretch}{1.12}
\setlength{\tabcolsep}{5pt}
\begin{tabular}{lccccc}
\hline\hline
State & $\sigma_{xx}/\tau$ & $\sigma_{yy}/\tau$ & $\sigma_{zz}/\tau$
& $\sigma_{zz}/\sigma_{xx}$ & $\sigma_{zz}/\sigma_{yy}$ \\
\hline
AM   & 494 & 671 & 1823 & 3.69 & 2.72 \\
AFM1 & 242 & 193 & 277  & 1.15 & 1.43 \\
AFM2 & 123 & 92  & 127  & 1.03 & 1.39 \\
\hline\hline
\end{tabular}
\end{table}

\bibliography{Ref_tidy}

@article{chen_large_2026,
	title        = {Large anisotropic magnetoresistance in $\ensuremath{\alpha}$-MnTe induced by strain},
	author       = {Chen, Bao-Feng and Yu, Jie-Xiang and Yin, Gen},
	year         = 2026,
	month        = {Feb},
	journal      = {Phys. Rev. B},
	publisher    = {American Physical Society},
	volume       = 113,
	pages        = {054412},
	doi          = {10.1103/v39q-ml7x},
	url          = {https://link.aps.org/doi/10.1103/v39q-ml7x},
	issue        = 5,
	numpages     = 6
}

@article{PhysRevX.12.031042,
	title        = {Beyond Conventional Ferromagnetism and Antiferromagnetism: A Phase with Nonrelativistic Spin and Crystal Rotation Symmetry},
	author       = {\ifmmode \check{S}\else \v{S}\fi{}mejkal, Libor and Sinova, Jairo and Jungwirth, Tomas},
	year         = 2022,
	month        = {Sep},
	journal      = {Phys. Rev. X},
	publisher    = {American Physical Society},
	volume       = 12,
	pages        = {031042},
	doi          = {10.1103/PhysRevX.12.031042},
	url          = {https://link.aps.org/doi/10.1103/PhysRevX.12.031042},
	issue        = 3,
	numpages     = 16
}

@article{PhysRevX.12.040501,
	title        = {Emerging Research Landscape of Altermagnetism},
	author       = {\ifmmode \check{S}\else \v{S}\fi{}mejkal, Libor and Sinova, Jairo and Jungwirth, Tomas},
	year         = 2022,
	month        = {Dec},
	journal      = {Phys. Rev. X},
	publisher    = {American Physical Society},
	volume       = 12,
	pages        = {040501},
	doi          = {10.1103/PhysRevX.12.040501},
	url          = {https://link.aps.org/doi/10.1103/PhysRevX.12.040501},
	issue        = 4,
	numpages     = 27
}

@article{Jungwirth2016,
	title        = {Antiferromagnetic spintronics},
	author       = {Jungwirth,  T. and Marti,  X. and Wadley,  P. and Wunderlich,  J.},
	year         = 2016,
	month        = mar,
	journal      = {Nature Nanotechnology},
	publisher    = {Springer Science and Business Media LLC},
	volume       = 11,
	number       = 3,
	pages        = {231–241},
	doi          = {10.1038/nnano.2016.18},
	issn         = {1748-3395},
	url          = {http://dx.doi.org/10.1038/nnano.2016.18}
}

@article{RevModPhys.90.015005,
	title        = {Antiferromagnetic spintronics},
	author       = {Baltz, V. and Manchon, A. and Tsoi, M. and Moriyama, T. and Ono, T. and Tserkovnyak, Y.},
	year         = 2018,
	month        = {Feb},
	journal      = {Rev. Mod. Phys.},
	publisher    = {American Physical Society},
	volume       = 90,
	pages        = {015005},
	doi          = {10.1103/RevModPhys.90.015005},
	url          = {https://link.aps.org/doi/10.1103/RevModPhys.90.015005},
	issue        = 1,
	numpages     = 57
}

@article{mejkal2022,
	title        = {Anomalous Hall antiferromagnets},
	author       = {Šmejkal,  Libor and MacDonald,  Allan H. and Sinova,  Jairo and Nakatsuji,  Satoru and Jungwirth,  Tomas},
	year         = 2022,
	month        = mar,
	journal      = {Nature Reviews Materials},
	publisher    = {Springer Science and Business Media LLC},
	volume       = 7,
	number       = 6,
	pages        = {482–496},
	doi          = {10.1038/s41578-022-00430-3},
	issn         = {2058-8437},
	url          = {http://dx.doi.org/10.1038/s41578-022-00430-3}
}

@article{Reimers2024,
	title        = {Direct observation of altermagnetic band splitting in {CrSb} thin films},
	author       = {Reimers,  Sonka and Odenbreit,  Lukas and Šmejkal,  Libor and Strocov,  Vladimir N. and Constantinou,  Procopios and Hellenes,  Anna B. and Jaeschke Ubiergo,  Rodrigo and Campos,  Warlley H. and Bharadwaj,  Venkata K. and Chakraborty,  Atasi and Denneulin,  Thibaud and Shi,  Wen and Dunin-Borkowski,  Rafal E. and Das,  Suvadip and Kl\"{a}ui,  Mathias and Sinova,  Jairo and Jourdan,  Martin},
	year         = 2024,
	month        = mar,
	journal      = {Nature Communications},
	publisher    = {Springer Science and Business Media LLC},
	volume       = 15,
	number       = 1,
	doi          = {10.1038/s41467-024-46476-5},
	issn         = {2041-1723},
	url          = {http://dx.doi.org/10.1038/s41467-024-46476-5}
}

@article{Fedchenko2024,
	title        = {Observation of time-reversal symmetry breaking in the band structure of altermagnetic RuO 2},
	author       = {Fedchenko,  Olena and Minár,  Jan and Akashdeep,  Akashdeep and D’Souza,  Sunil Wilfred and Vasilyev,  Dmitry and Tkach,  Olena and Odenbreit,  Lukas and Nguyen,  Quynh and Kutnyakhov,  Dmytro and Wind,  Nils and Wenthaus,  Lukas and Scholz,  Markus and Rossnagel,  Kai and Hoesch,  Moritz and Aeschlimann,  Martin and Stadtm\"{u}ller,  Benjamin and Kl\"{a}ui,  Mathias and Sch\"{o}nhense,  Gerd and Jungwirth,  Tomas and Hellenes,  Anna Birk and Jakob,  Gerhard and Šmejkal,  Libor and Sinova,  Jairo and Elmers,  Hans-Joachim},
	year         = 2024,
	month        = feb,
	journal      = {Science Advances},
	publisher    = {American Association for the Advancement of Science (AAAS)},
	volume       = 10,
	number       = 5,
	doi          = {10.1126/sciadv.adj4883},
	issn         = {2375-2548},
	url          = {http://dx.doi.org/10.1126/sciadv.adj4883}
}

@article{Feng2022,
	title        = {An anomalous Hall effect in altermagnetic ruthenium dioxide},
	author       = {Feng,  Zexin and Zhou,  Xiaorong and Šmejkal,  Libor and Wu,  Lei and Zhu,  Zengwei and Guo,  Huixin and González-Hernández,  Rafael and Wang,  Xiaoning and Yan,  Han and Qin,  Peixin and Zhang,  Xin and Wu,  Haojiang and Chen,  Hongyu and Meng,  Ziang and Liu,  Li and Xia,  Zhengcai and Sinova,  Jairo and Jungwirth,  Tomáš and Liu,  Zhiqi},
	year         = 2022,
	month        = nov,
	journal      = {Nature Electronics},
	publisher    = {Springer Science and Business Media LLC},
	volume       = 5,
	number       = 11,
	pages        = {735–743},
	doi          = {10.1038/s41928-022-00866-z},
	issn         = {2520-1131},
	url          = {http://dx.doi.org/10.1038/s41928-022-00866-z}
}

@article{PhysRevMaterials.9.064403,
	title        = {High-throughput screening of altermagnetic materials},
	author       = {Bhattarai, Romakanta and Minch, Peter and Rhone, Trevor David},
	year         = 2025,
	month        = {Jun},
	journal      = {Phys. Rev. Mater.},
	publisher    = {American Physical Society},
	volume       = 9,
	pages        = {064403},
	doi          = {10.1103/PhysRevMaterials.9.064403},
	url          = {https://link.aps.org/doi/10.1103/PhysRevMaterials.9.064403},
	issue        = 6,
	numpages     = 7
}

@article{Rondinelli2011,
	title        = {Structure and Properties of Functional Oxide Thin Films: Insights From Electronic‐Structure Calculations},
	author       = {Rondinelli,  James M. and Spaldin,  Nicola A.},
	year         = 2011,
	month        = jul,
	journal      = {Advanced Materials},
	publisher    = {Wiley},
	volume       = 23,
	number       = 30,
	pages        = {3363–3381},
	doi          = {10.1002/adma.201101152},
	issn         = {1521-4095},
	url          = {http://dx.doi.org/10.1002/adma.201101152}
}

@article{PhysRevB.72.045103,
	title        = {Kondo insulator description of spin state transition in $\mathrm{Fe}{\mathrm{Sb}}_{2}$},
	author       = {Petrovic, C. and Lee, Y. and Vogt, T. and Lazarov, N. Dj. and Bud'ko, S. L. and Canfield, P. C.},
	year         = 2005,
	month        = {Jul},
	journal      = {Phys. Rev. B},
	publisher    = {American Physical Society},
	volume       = 72,
	pages        = {045103},
	doi          = {10.1103/PhysRevB.72.045103},
	url          = {https://link.aps.org/doi/10.1103/PhysRevB.72.045103},
	issue        = 4,
	numpages     = 7
}

@article{Bentien2007,
	title        = {Colossal Seebeck coefficient in strongly correlated semiconductor {FeSb$_2$}},
	author       = {Bentien,  A. and Johnsen,  S. and Madsen,  G. K. H. and Iversen,  B. B. and Steglich,  F.},
	year         = 2007,
	month        = sep,
	journal      = {Europhysics Letters (EPL)},
	publisher    = {IOP Publishing},
	volume       = 80,
	number       = 1,
	pages        = 17008,
	doi          = {10.1209/0295-5075/80/17008},
	issn         = {1286-4854},
	url          = {http://dx.doi.org/10.1209/0295-5075/80/17008}
}

@article{PhysRevLett.114.236603,
	title        = {Unified Picture for the Colossal Thermopower Compound ${\mathrm{FeSb}}_{2}$},
	author       = {Battiato, M. and Tomczak, J. M. and Zhong, Z. and Held, K.},
	year         = 2015,
	month        = {Jun},
	journal      = {Phys. Rev. Lett.},
	publisher    = {American Physical Society},
	volume       = 114,
	pages        = 236603,
	doi          = {10.1103/PhysRevLett.114.236603},
	url          = {https://link.aps.org/doi/10.1103/PhysRevLett.114.236603},
	issue        = 23,
	numpages     = 5
}

@article{Takahashi2016,
	title        = {Colossal Seebeck effect enhanced by quasi-ballistic phonons dragging massive electrons in {FeSb$_2$}},
	author       = {Takahashi,  H. and Okazaki,  R. and Ishiwata,  S. and Taniguchi,  H. and Okutani,  A. and Hagiwara,  M. and Terasaki,  I.},
	year         = 2016,
	month        = sep,
	journal      = {Nature Communications},
	publisher    = {Springer Science and Business Media LLC},
	volume       = 7,
	number       = 1,
	doi          = {10.1038/ncomms12732},
	issn         = {2041-1723},
	url          = {http://dx.doi.org/10.1038/ncomms12732}
}

@article{Xu2020,
	title        = {Metallic surface states in a correlated d-electron topological Kondo insulator candidate {FeSb$_2$}},
	author       = {Xu,  Ke-Jun and Chen,  Su-Di and He,  Yu and He,  Junfeng and Tang,  Shujie and Jia,  Chunjing and Yue Ma,  Eric and Mo,  Sung-Kwan and Lu,  Donghui and Hashimoto,  Makoto and Devereaux,  Thomas P. and Shen,  Zhi-Xun},
	year         = 2020,
	month        = jun,
	journal      = {Proceedings of the National Academy of Sciences},
	publisher    = {Proceedings of the National Academy of Sciences},
	volume       = 117,
	number       = 27,
	pages        = {15409–15413},
	doi          = {10.1073/pnas.2002361117},
	issn         = {1091-6490},
	url          = {http://dx.doi.org/10.1073/pnas.2002361117}
}

@article{Li2024,
	title        = {Spectroscopic evidence of spin-state excitation in d-electron correlated semiconductor {FeSb$_2$}},
	author       = {Li,  Huayao and Wang,  Guohua and Ding,  Ning and Ren,  Quan and Zhao,  Gan and Lin,  Wenting and Yang,  Jinchuan and Yan,  Wensheng and Li,  Qian and Yang,  Run and Yuan,  Shijun and Denlinger,  Jonathan D. and Wang,  Zhenxing and Zhang,  Xiaoqian and Wray,  L. Andrew and Dong,  Shuai and Qian,  Dong and Miao,  Lin},
	year         = 2024,
	month        = jul,
	journal      = {Proceedings of the National Academy of Sciences},
	publisher    = {Proceedings of the National Academy of Sciences},
	volume       = 121,
	number       = 28,
	doi          = {10.1073/pnas.2321193121},
	issn         = {1091-6490},
	url          = {http://dx.doi.org/10.1073/pnas.2321193121}
}

@article{Mazin2021,
	title        = {Prediction of unconventional magnetism in doped {FeSb$_2$}},
	author       = {Mazin,  Igor I. and Koepernik,  Klaus and Johannes,  Michelle D. and González-Hernández,  Rafael and Šmejkal,  Libor},
	year         = 2021,
	month        = oct,
	journal      = {Proceedings of the National Academy of Sciences},
	publisher    = {Proceedings of the National Academy of Sciences},
	volume       = 118,
	number       = 42,
	doi          = {10.1073/pnas.2108924118},
	issn         = {1091-6490},
	url          = {http://dx.doi.org/10.1073/pnas.2108924118}
}

@article{PhysRevX.12.040002,
	title        = {Editorial: Altermagnetism---A New Punch Line of Fundamental Magnetism},
	author       = {Mazin, Igor},
	year         = 2022,
	month        = {Dec},
	journal      = {Phys. Rev. X},
	publisher    = {American Physical Society},
	volume       = 12,
	pages        = {040002},
	doi          = {10.1103/PhysRevX.12.040002},
	url          = {https://link.aps.org/doi/10.1103/PhysRevX.12.040002},
	collaboration = {The PRX Editors},
	issue        = 4,
	numpages     = 3
}

@article{doi:10.7566/JPSJ.88.123702,
	title        = {Momentum-Dependent Spin Splitting by Collinear Antiferromagnetic Ordering},
	author       = {Hayami ,Satoru and Yanagi ,Yuki and Kusunose ,Hiroaki},
	year         = 2019,
	journal      = {Journal of the Physical Society of Japan},
	volume       = 88,
	number       = 12,
	pages        = 123702,
	doi          = {10.7566/JPSJ.88.123702},
	url          = {https://doi.org/10.7566/JPSJ.88.123702}
}

@article{PhysRevB.102.014422,
	title        = {Giant momentum-dependent spin splitting in centrosymmetric low-$Z$ antiferromagnets},
	author       = {Yuan, Lin-Ding and Wang, Zhi and Luo, Jun-Wei and {R}ashba, Emmanuel I. and Zunger, Alex},
	year         = 2020,
	month        = {Jul},
	journal      = {Phys. Rev. B},
	publisher    = {American Physical Society},
	volume       = 102,
	pages        = {014422},
	doi          = {10.1103/PhysRevB.102.014422},
	url          = {https://link.aps.org/doi/10.1103/PhysRevB.102.014422},
	issue        = 1,
	numpages     = 13
}

@article{PhysRevLett.132.036702,
	title        = {Broken {K}ramers Degeneracy in Altermagnetic {M}n{T}e},
	author       = {Lee, Suyoung and Lee, Sangjae and Jung, Saegyeol and Jung, Jiwon and Kim, Donghan and Lee, Yeonjae and Seok, Byeongjun and Kim, Jaeyoung and Park, Byeong Gyu and \ifmmode \check{S}\else \v{S}\fi{}mejkal, Libor and Kang, Chang-Jong and Kim, Changyoung},
	year         = 2024,
	month        = {Jan},
	journal      = {Phys. Rev. Lett.},
	publisher    = {American Physical Society},
	volume       = 132,
	pages        = {036702},
	doi          = {10.1103/PhysRevLett.132.036702},
	url          = {https://link.aps.org/doi/10.1103/PhysRevLett.132.036702},
	issue        = 3,
	numpages     = 7
}

@article{PhysRevMaterials.5.014409,
	title        = {Prediction of low-Z collinear and noncollinear antiferromagnetic compounds having momentum-dependent spin splitting even without spin-orbit coupling},
	author       = {Yuan, Lin-Ding and Wang, Zhi and Luo, Jun-Wei and Zunger, Alex},
	year         = 2021,
	month        = {Jan},
	journal      = {Phys. Rev. Mater.},
	publisher    = {American Physical Society},
	volume       = 5,
	pages        = {014409},
	doi          = {10.1103/PhysRevMaterials.5.014409},
	url          = {https://link.aps.org/doi/10.1103/PhysRevMaterials.5.014409},
	issue        = 1,
	numpages     = 24
}

@article{PhysRevB.99.184432,
	title        = {Antiferromagnetism in {R}u{O}$_2$ as $d$-wave Pomeranchuk instability},
	author       = {Ahn, Kyo-Hoon and Hariki, Atsushi and Lee, Kwan-Woo and Kune\ifmmode \check{s}\else \v{s}\fi{}, Jan},
	year         = 2019,
	month        = {May},
	journal      = {Phys. Rev. B},
	publisher    = {American Physical Society},
	volume       = 99,
	pages        = 184432,
	doi          = {10.1103/PhysRevB.99.184432},
	url          = {https://link.aps.org/doi/10.1103/PhysRevB.99.184432},
	issue        = 18,
	numpages     = 5
}

@article{PhysRevB.109.144421,
	title        = {Strain-induced phase transition from antiferromagnet to altermagnet},
	author       = {Chakraborty, Atasi and Gonz\'alez Hern\'andez, Rafael and \ifmmode \check{S}\else \v{S}\fi{}mejkal, Libor and Sinova, Jairo},
	year         = 2024,
	month        = {Apr},
	journal      = {Phys. Rev. B},
	publisher    = {American Physical Society},
	volume       = 109,
	pages        = 144421,
	doi          = {10.1103/PhysRevB.109.144421},
	url          = {https://link.aps.org/doi/10.1103/PhysRevB.109.144421},
	issue        = 14,
	numpages     = 12
}

@article{he2025altermagnetismttprimedeltafermihubbardmodel,
  title = {Altermagnetism and beyond in the $t\text{\ensuremath{-}}{t}^{\ensuremath{'}}\text{\ensuremath{-}}\ensuremath{\delta}$ Fermi-Hubbard model},
  author = {He, Saisai and Zhao, Jize and Luo, Hong-Gang and Hu, Shijie},
  journal = {Phys. Rev. B},
  volume = {112},
  issue = {3},
  pages = {035108},
  numpages = {13},
  year = {2025},
  month = {Jul},
  publisher = {American Physical Society},
  doi = {10.1103/4mv8-tb66},
  url = {https://link.aps.org/doi/10.1103/4mv8-tb66}
}

@article{PhysRevX.12.011028,
	title        = {Giant and Tunneling Magnetoresistance in Unconventional Collinear Antiferromagnets with Nonrelativistic Spin-Momentum Coupling},
	author       = {\ifmmode \check{S}\else \v{S}\fi{}mejkal, Libor and Hellenes, Anna Birk and Gonz\'alez-Hern\'andez, Rafael and Sinova, Jairo and Jungwirth, Tomas},
	year         = 2022,
	month        = {Feb},
	journal      = {Phys. Rev. X},
	publisher    = {American Physical Society},
	volume       = 12,
	pages        = {011028},
	doi          = {10.1103/PhysRevX.12.011028},
	url          = {https://link.aps.org/doi/10.1103/PhysRevX.12.011028},
	issue        = 1,
	numpages     = 11
}

@article{Krempaský2024,
	title        = {Altermagnetic lifting of {K}ramers spin degeneracy},
	author       = {Krempask{\'y}, J. and {\v{S}}mejkal, L. and D'Souza, S. W. and Hajlaoui, M. and Springholz, G. and Uhl{\'i}{\v{r}}ov{\'a}, K. and Alarab, F. and Constantinou, P. C. and Strocov, V. and Usanov, D. and others},
	year         = 2024,
	month        = {Feb},
	day          = {01},
	journal      = {Nature},
	volume       = 626,
	number       = 7999,
	pages        = {517--522},
	doi          = {10.1038/s41586-023-06907-7},
	issn         = {1476-4687},
	url          = {https://doi.org/10.1038/s41586-023-06907-7}
}

@article{PhysRevLett.126.127701,
	title        = {Efficient Electrical Spin Splitter Based on Nonrelativistic Collinear Antiferromagnetism},
	author       = {Gonz\'alez-Hern\'andez, Rafael and \ifmmode \check{S}\else \v{S}\fi{}mejkal, Libor and V\'yborn\'y, Karel and Yahagi, Yuta and Sinova, Jairo and Jungwirth, Tom\'a\ifmmode\check{s}\else\v{s}\fi{} and \ifmmode \check{Z}\else \v{Z}\fi{}elezn\'y, Jakub},
	year         = 2021,
	month        = {Mar},
	journal      = {Phys. Rev. Lett.},
	publisher    = {American Physical Society},
	volume       = 126,
	pages        = 127701,
	doi          = {10.1103/PhysRevLett.126.127701},
	url          = {https://link.aps.org/doi/10.1103/PhysRevLett.126.127701},
	issue        = 12,
	numpages     = 6
}

@misc{golub2025spinorientationelectriccurrent,
      title={Spin orientation by electric current in altermagnets}, 
      author={L. E. Golub and L. Šmejkal},
      year={2025},
      eprint={2503.12203},
      archivePrefix={arXiv},
      primaryClass={cond-mat.mes-hall},
      url={https://arxiv.org/abs/2503.12203}, 
}

@article{Yarmohammadi2025Spin,
	title        = {Spin polarization engineering in $d$-wave altermagnets},
	author       = {Yarmohammadi, Mohsen and Berritta, Marco and Bukov, Marin and \ifmmode \check{S}\else \v{S}\fi{}mejkal, Libor and Linder, Jacob and Oppeneer, Peter M.},
	year         = 2026,
	month        = {Feb},
	journal      = {Phys. Rev. B},
	publisher    = {American Physical Society},
	volume       = 113,
	pages        = {L060403},
	doi          = {10.1103/xt23-9pnv},
	url          = {https://link.aps.org/doi/10.1103/xt23-9pnv},
	issue        = 6,
	numpages     = 6
}

@article{Yarmohammadi2025Cavity,
	title        = {Cavity-Induced Coherent Magnetization and Polaritons in Altermagnets},
	author       = {Yarmohammadi, Mohsen and \ifmmode \check{S}\else \v{S}\fi{}mejkal, Libor and Freericks, James K.},
	year         = 2026,
	month        = {Apr},
	journal      = {Phys. Rev. Lett.},
	publisher    = {American Physical Society},
	volume       = 136,
	pages        = 146904,
	doi          = {10.1103/7bss-9yxb},
	url          = {https://link.aps.org/doi/10.1103/7bss-9yxb},
	issue        = 14,
	numpages     = 9
}

@article{Yarmohammadi2025SlowPhonon,
	title        = {Slow-phonon control of spin Edelstein effect in Rashba $d$-wave altermagnets},
	author       = {Yarmohammadi, Mohsen and Linder, Jacob and Freericks, James K.},
	year         = 2026,
	month        = {May},
	journal      = {Phys. Rev. B},
	publisher    = {American Physical Society},
	volume       = 113,
	pages        = 184403,
	doi          = {10.1103/4txy-gy8f},
	url          = {https://link.aps.org/doi/10.1103/4txy-gy8f},
	issue        = 18,
	numpages     = 14
}

@article{Milivojevi2024,
	title        = {Interplay of altermagnetism and weak ferromagnetism in two-dimensional {RuF$_4$}},
	author       = {Milivojević,  Marko and Orozović,  Marko and Picozzi,  Silvia and Gmitra,  Martin and Stavrić,  Srđan},
	year         = 2024,
	month        = may,
	journal      = {2D Materials},
	publisher    = {IOP Publishing},
	volume       = 11,
	number       = 3,
	pages        = {035025},
	doi          = {10.1088/2053-1583/ad4c73},
	issn         = {2053-1583},
	url          = {http://dx.doi.org/10.1088/2053-1583/ad4c73}
}

@article{Aoyama2024MnTe,
	title        = {Piezomagnetic properties in altermagnetic {MnTe}},
	author       = {Aoyama, Takuya and Ohgushi, Kenya},
	year         = 2024,
	month        = {Apr},
	journal      = {Phys. Rev. Mater.},
	publisher    = {American Physical Society},
	volume       = 8,
	pages        = {L041402},
	doi          = {10.1103/PhysRevMaterials.8.L041402},
	url          = {https://link.aps.org/doi/10.1103/PhysRevMaterials.8.L041402},
	issue        = 4,
	numpages     = 5
}

@article{Yershov2024,
	title        = {Fluctuation-induced piezomagnetism in local moment altermagnets},
	author       = {Yershov, Kostiantyn V. and Kravchuk, Volodymyr P. and Daghofer, Maria and van den Brink, Jeroen},
	year         = 2024,
	month        = {Oct},
	journal      = {Phys. Rev. B},
	publisher    = {American Physical Society},
	volume       = 110,
	pages        = 144421,
	doi          = {10.1103/PhysRevB.110.144421},
	url          = {https://link.aps.org/doi/10.1103/PhysRevB.110.144421},
	issue        = 14,
	numpages     = 11
}

@article{Ogawa2025,
	title        = {Nonlinear Piezomagnetic Effects in g-wave Altermagnets},
	author       = {Ogawa,  Yuuki and Hayami,  Satoru},
	year         = 2025,
	month        = jun,
	journal      = {Journal of the Physical Society of Japan},
	publisher    = {Physical Society of Japan},
	volume       = 94,
	number       = 6,
	doi          = {10.7566/jpsj.94.063704},
	issn         = {1347-4073},
	url          = {http://dx.doi.org/10.7566/JPSJ.94.063704}
}

@article{Hodt2024,
	title        = {Interface-induced magnetization in altermagnets and antiferromagnets},
	author       = {Hodt, Erik Wegner and Sukhachov, Pavlo and Linder, Jacob},
	year         = 2024,
	month        = {Aug},
	journal      = {Phys. Rev. B},
	publisher    = {American Physical Society},
	volume       = 110,
	pages        = {054446},
	doi          = {10.1103/PhysRevB.110.054446},
	url          = {https://link.aps.org/doi/10.1103/PhysRevB.110.054446},
	issue        = 5,
	numpages     = 17
}

@article{Kresse1996PRB,
	title        = {Efficient iterative schemes for \textit{ab initio} total-energy calculations using a plane-wave basis set},
	author       = {Kresse, G. and Furthm{\"u}ller, J.},
	year         = 1996,
	journal      = {Phys. Rev. B},
	volume       = 54,
	pages        = {11169--11186},
	doi          = {10.1103/PhysRevB.54.11169},
	url          = {https://doi.org/10.1103/PhysRevB.54.11169}
}

@article{Kresse1996CMS,
	title        = {Efficiency of \textit{ab-initio} total energy calculations for metals and semiconductors using a plane-wave basis set},
	author       = {Kresse, G. and Furthm{\"u}ller, J.},
	year         = 1996,
	journal      = {Comput. Mater. Sci.},
	volume       = 6,
	number       = 1,
	pages        = {15--50},
	doi          = {10.1016/0927-0256(96)00008-0},
	url          = {https://doi.org/10.1016/0927-0256(96)00008-0}
}

@article{Blochl1994,
	title        = {Projector augmented-wave method},
	author       = {Bl{\"o}chl, P. E.},
	year         = 1994,
	journal      = {Phys. Rev. B},
	volume       = 50,
	pages        = {17953--17979},
	doi          = {10.1103/PhysRevB.50.17953},
	url          = {https://doi.org/10.1103/PhysRevB.50.17953}
}

@article{KresseJoubert1999,
	title        = {From ultrasoft pseudopotentials to the projector augmented-wave method},
	author       = {Kresse, G. and Joubert, D.},
	year         = 1999,
	journal      = {Phys. Rev. B},
	volume       = 59,
	pages        = {1758--1775},
	doi          = {10.1103/PhysRevB.59.1758},
	url          = {https://doi.org/10.1103/PhysRevB.59.1758}
}

@article{PBE1996,
	title        = {Generalized Gradient Approximation Made Simple},
	author       = {Perdew, John P. and Burke, Kieron and Ernzerhof, Matthias},
	year         = 1996,
	journal      = {Phys. Rev. Lett.},
	volume       = 77,
	pages        = {3865--3868},
	doi          = {10.1103/PhysRevLett.77.3865},
	url          = {https://doi.org/10.1103/PhysRevLett.77.3865}
}

@article{MonkhorstPack1976,
	title        = {Special points for Brillouin-zone integrations},
	author       = {Monkhorst, Hendrik J. and Pack, James D.},
	year         = 1976,
	journal      = {Phys. Rev. B},
	volume       = 13,
	pages        = {5188--5192},
	doi          = {10.1103/PhysRevB.13.5188},
	url          = {https://doi.org/10.1103/PhysRevB.13.5188}
}

@article{MarzariVanderbilt1997,
	title        = {Maximally localized generalized Wannier functions for composite energy bands},
	author       = {Marzari, Nicola and Vanderbilt, David},
	year         = 1997,
	journal      = {Phys. Rev. B},
	volume       = 56,
	pages        = {12847--12865},
	doi          = {10.1103/PhysRevB.56.12847},
	url          = {https://doi.org/10.1103/PhysRevB.56.12847}
}

@article{Souza2001,
	title        = {Maximally localized Wannier functions for entangled energy bands},
	author       = {Souza, Ivo and Marzari, Nicola and Vanderbilt, David},
	year         = 2001,
	journal      = {Phys. Rev. B},
	volume       = 65,
	pages        = {035109},
	doi          = {10.1103/PhysRevB.65.035109},
	url          = {https://doi.org/10.1103/PhysRevB.65.035109}
}

@article{Mostofi2008,
	title        = {wannier90: A tool for obtaining maximally-localised Wannier functions},
	author       = {Mostofi, Arash A. and Yates, Jonathan R. and Lee, Young-Su and Souza, Ivo and Vanderbilt, David and Marzari, Nicola},
	year         = 2008,
	journal      = {Comput. Phys. Commun.},
	volume       = 178,
	number       = 9,
	pages        = {685--699},
	doi          = {10.1016/j.cpc.2007.11.016},
	url          = {https://doi.org/10.1016/j.cpc.2007.11.016}
}

@article{Mostofi2014,
	title        = {An updated version of wannier90: A tool for obtaining maximally-localised Wannier functions},
	author       = {Mostofi, Arash A. and Yates, Jonathan R. and Pizzi, Giovanni and Lee, Young-Su and Souza, Ivo and Vanderbilt, David and Marzari, Nicola},
	year         = 2014,
	journal      = {Comput. Phys. Commun.},
	volume       = 185,
	number       = 8,
	pages        = {2309--2310},
	doi          = {10.1016/j.cpc.2014.05.003},
	url          = {https://doi.org/10.1016/j.cpc.2014.05.003}
}

@article{Madsen2006BoltzTraP,
	title        = {BoltzTraP. A code for calculating band-structure dependent quantities},
	author       = {Madsen, Georg K. H. and Singh, David J.},
	year         = 2006,
	journal      = {Comput. Phys. Commun.},
	volume       = 175,
	number       = 1,
	pages        = {67--71},
	doi          = {10.1016/j.cpc.2006.03.007}
}

@book{ZimanBook,
	title        = {Electrons and Phonons: The Theory of Transport Phenomena in Solids},
	author       = {Ziman, J. M.},
	year         = 1960,
	publisher    = {Oxford University Press}
}

@article{Pizzi2020,
	title        = {Wannier90 as a community code: new features and applications},
	author       = {Pizzi, Giovanni and Vitale, Valerio and Arita, Ryotaro and Bl{\"u}gel, Stefan and Freimuth, Frank and G{\'e}ranton, Guillaume and Gibertini, Marco and Gresch, Dominik and Johnson, Charles and Koretsune, Takashi and Ib{\'a}{\~n}ez-Azpiroz, Julen and Lee, Hyungjun and Lihm, Jae-Mo and Marchand, Daniel and Marrazzo, Antimo and Mokrousov, Yuriy and Mustafa, Jamal I. and Nohara, Yoshiro and Nomura, Yusuke and Paulatto, Lorenzo and Ponc{\'e}, Samuel and Ponweiser, Thomas and Qiao, Junfeng and Th{\"o}le, Florian and Tsirkin, Stepan S. and Wierzbowska, Ma{\l}gorzata and Marzari, Nicola and Vanderbilt, David and Souza, Ivo and Mostofi, Arash A. and Yates, Jonathan R.},
	year         = 2020,
	journal      = {J. Phys.: Condens. Matter},
	volume       = 32,
	number       = 16,
	pages        = 165902,
	doi          = {10.1088/1361-648X/ab51ff},
	url          = {https://doi.org/10.1088/1361-648X/ab51ff}
}

@article{Yu2021PRB,
	title        = {Discrete quantum geometry and intrinsic spin Hall effect},
	author       = {Yu, Jie-Xiang and Zang, Jiadong and Lake, Roger K. and Zhang, Yi and Yin, Gen},
	year         = 2021,
	journal      = {Physical Review B},
	volume       = 104,
	number       = 18,
	pages        = 184408,
	doi          = {10.1103/PhysRevB.104.184408},
	url          = {https://doi.org/10.1103/PhysRevB.104.184408}
}

@misc{Shawon2026CrDopedFeSb2,
	title        = {Evidence for altermagnetic order in Cr-doped {FeSb$_2$}},
	author       = {Shawon, A K M Ashiquzzaman and Downey, Eoghan and Smolenski, Shane and Hicken, Thomas J. and Henderson, Amir and Xu, Mingyu and Musall, Trisha and Lopes Sabainsk, Rafael and Zhu, Yuan and Xie, Weiwei and Gati, Elena and Li, Lu and Guguchia, Zurab and Jo, Na Hyun},
	year         = 2026,
	eprint       = {2605.01088},
	archiveprefix = {arXiv},
	primaryclass = {cond-mat.mtrl-sci}
}

@article{Tomczak2010ThermopowerFeSb2,
	title        = {Thermopower of correlated semiconductors: Application to ${\text{FeAs}}_{2}$ and ${\text{FeSb}}_{2}$},
	author       = {Tomczak, Jan M. and Haule, K. and Miyake, T. and Georges, A. and Kotliar, G.},
	year         = 2010,
	month        = {Aug},
	journal      = {Phys. Rev. B},
	publisher    = {American Physical Society},
	volume       = 82,
	pages        = {085104},
	doi          = {10.1103/PhysRevB.82.085104},
	url          = {https://link.aps.org/doi/10.1103/PhysRevB.82.085104},
	issue        = 8,
	numpages     = 13
}

@article{Kang2018FeSb2Analogs,
	title        = {Study for material analogs of ${\mathrm{FeSb}}_{2}$: Material design for thermoelectric materials},
	author       = {Kang, Chang-Jong and Kotliar, Gabriel},
	year         = 2018,
	month        = {Mar},
	journal      = {Phys. Rev. Mater.},
	publisher    = {American Physical Society},
	volume       = 2,
	pages        = {034604},
	doi          = {10.1103/PhysRevMaterials.2.034604},
	url          = {https://link.aps.org/doi/10.1103/PhysRevMaterials.2.034604},
	issue        = 3,
	numpages     = 11
}

@article{Lukoyanov2006FeSb2,
	title        = {The semiconductor-to-ferromagnetic-metal transition in FeSb2},
	author       = {Lukoyanov, A. V. and Mazurenko, V. V. and Anisimov, V. I. and Sigrist, M. and Rice, T. M.},
	year         = 2006,
	journal      = {The European Physical Journal B},
	volume       = 53,
	pages        = {205--207},
	doi          = {10.1140/epjb/e2006-00361-0}
}

@article{Kuhn2013CrSb2FeSb2,
	title        = {Electronic structure and magnetic properties of CrSb${}_{2}$ and FeSb${}_{2}$ investigated via ab initio calculations},
	author       = {Kuhn, G. and Mankovsky, S. and Ebert, H. and Regus, M. and Bensch, W.},
	year         = 2013,
	month        = {Feb},
	journal      = {Phys. Rev. B},
	publisher    = {American Physical Society},
	volume       = 87,
	pages        = {085113},
	doi          = {10.1103/PhysRevB.87.085113},
	url          = {https://link.aps.org/doi/10.1103/PhysRevB.87.085113},
	issue        = 8,
	numpages     = 8
}

@article{PhysRevB.111.075141,
	title        = {Electronic structure of the altermagnet candidate ${\mathrm{FeSb}}_{2}$: High-field torque magnetometry and density functional theory studies},
	author       = {Phillips, Cole and Pokharel, Ganesh and Shtefiienko, Kyryl and Bhandari, Shalika R. and Graf, David E. and Rai, D. P. and Shrestha, Keshav},
	year         = 2025,
	month        = {Feb},
	journal      = {Phys. Rev. B},
	publisher    = {American Physical Society},
	volume       = 111,
	pages        = {075141},
	doi          = {10.1103/PhysRevB.111.075141},
	url          = {https://link.aps.org/doi/10.1103/PhysRevB.111.075141},
	issue        = 7,
	numpages     = 9
}

@article{Yarmohammadi2026FloquetAMR,
  title = {Floquet-induced anisotropic magnetoresistance and anomalous Hall effect in 2D d-wave altermagnets with Rashba spin-orbit coupling},
  author = {Yarmohammadi, Mohsen and Gunnink, Pieter M. and Sinova, Jairo and Freericks, James K.},
  journal = {arXiv preprint arXiv:2606.21281},
  year = {2026},
  eprint = {2606.21281},
  archivePrefix = {arXiv},
  primaryClass = {cond-mat.mtrl-sci}
}
\end{document}